\documentclass[structabstract]{aa} 
\usepackage[utf8]{inputenc}
\usepackage{txfonts} 
\usepackage{graphicx} 
\usepackage{natbib} 
\usepackage{url} 
\usepackage{float}
\usepackage{graphicx}   % Including figure files
\usepackage{amsmath}    % Advanced maths commands
\usepackage{amssymb}    % Extra maths symbols
\usepackage[dvipsnames]{xcolor}

\usepackage{comment}

\usepackage[colorlinks = true,
            linkcolor = blue,
            urlcolor  = blue,
            citecolor = blue,
            anchorcolor = blue]{hyperref}

\usepackage{ulem}

\usepackage{longtable}

\usepackage{threeparttable}

\usepackage{colortbl}

\usepackage{subcaption}
\usepackage{mwe}

\usepackage{enumitem}

\usepackage{pifont}
\usepackage{dirtytalk}

\newcommand{\be}{\begin{equation}}
\newcommand{\ee}{\end{equation}}

\newcommand{\mstar}{M_{\star}}

\newcommand{\mh}{M_{\rm BH}}

\def\msun{\, \mathrm{M}_{\hbox{$\odot$}}}

\newcommand{\tbb}{T_{\rm BB}}
\newcommand{\rbb}{R_{\rm BB}}

\begin{document}

   \title{AT~2020wey and the class of faint and fast tidal disruption events}

    \titlerunning{AT~2020wey and the class of faint and fast Tidal Disruption Events}
%   \subtitle{I. Overviewing the $\kappa$-mechanism}

   \author{P. Charalampopoulos\inst{1,2}\fnmsep\thanks{Contact e-mail: \href{mailto:pngchr@space.dtu.dk}{pngchr@space.dtu.dk}}\href{https://orcid.org/0000-0002-7706-5668}\
          \and
          M. Pursiainen\inst{1}\href{https://orcid.org/0000-0003-4663-4300}
          \and
          G. Leloudas \inst{1}\href{https://orcid.org/0000-0002-8597-0756} 
          \and
          I. Arcavi \inst{3,4}\href{https://orcid.org/0000-0001-7090-4898}
          \and
          M. Newsome\inst{5,6}
          \and
          S. Schulze\inst{7}\href{https://orcid.org/0000-0001-6797-1889}
          \and
          J. Burke\inst{5,6}\href{https://orcid.org/0000-0003-0035-6659}
          \and
          M. Nicholl\inst{8,9}\href{https://orcid.org/0000-0002-2555-3192}
            \
          }

   \institute{DTU Space, National Space Institute, Technical University of Denmark, Elektrovej 327, DK-2800 Kgs. Lyngby, Denmark
            %  \email{wuchterl@amok.ast.univie.ac.at}
         \and
         Department of Physics and Astronomy, University of Turku, Vesilinnantie 5, FI-20500, Finland
         \and
         The School of Physics and Astronomy, Tel Aviv University, Tel Aviv 69978, Israel
         \and
         CIFAR Azrieli Global Scholars program, CIFAR, Toronto, Canada
         \and
         Las Cumbres Observatory, 6740 Cortona Dr, Suite 102, Goleta,
        CA 93117-5575, USA
        \and
        Department of Physics, University of California, Santa Barbara, CA
        93106-9530, USA
        \and
         The Oskar Klein Centre, Department of Astronomy, Stockholm University, AlbaNova, SE-10691 Stockholm, Sweden
         \and
         School of Physics and Astronomy, University of Birmingham, Birmingham B15 2TT, UK
         \and
         Institute for Gravitational Wave Astronomy, University of Birmingham, Birmingham B15 2TT, UK
             }

   \date{Received - ; accepted -}

% \abstract{}{}{}{}{} 
% 5 {} token are mandatory
 
%   \abstract
%   % context heading (optional)
%   % {} leave it empty if necessary  

%   % conclusions heading (optional), leave it empty if necessary 
%   {}

  \abstract
   {We present an analysis of the optical and ultraviolet properties of AT~2020wey, a faint and fast tidal disruption event (TDE) at 124.3 Mpc. The light curve of the object peaked at an absolute magnitude of $M_{g} = -17.45\pm0.08$~mag and a maximum bolometric luminosity of $L_{\rm peak}=(8.74\pm0.69)\times10^{42}$\,erg\,s$^{-1}$, making it comparable  to iPTF16fnl, the faintest TDE to date. The time from the last non-detection to the $g$-band peak is 23 $\pm$ 2 days, and the rise is well described by $L\propto t^{1.80\pm0.22}$. The decline of the bolometric light curve is described by a sharp exponential decay steeper than the canonical $t^{-5/3}$ power law, making AT~2020wey the fastest declining TDE to date. The multi-band light curve analysis shows first a slowly declining  blackbody temperature of $\tbb\sim20\,000$~K around the peak brightness followed by a gradual temperature increase. The blackbody photosphere is found to expand at a constant velocity ($\sim 1\,300$\,km\,s$^{-1}$) to a value of $\rbb\sim3.5\times10^{14}$~cm before contracting rapidly. Multi-wavelength fits to the light curve indicate a complete disruption of a star of $\mstar=0.11^{+0.05}_{-0.02} \msun$ by a black hole of $\mh=10^{6.46^{+0.09}_{-0.09}} \msun$. Our spectroscopic dataset reveals broad ($\sim10^{4}$\,km\,s$^{-1}$) Balmer and \ion{He}{II} 4686~\AA\, lines, with H$\alpha$ reaching its peak with a lag of $\sim8.2$ days compared to the continuum. 
   In contrast to previous faint and fast TDEs, there are no obvious Bowen fluorescence lines in the spectra of AT~2020wey. 
    There is a strong correlation between the \texttt{MOSFIT}-derived black hole masses of TDEs and their decline rate. However, AT~2020wey is an outlier in this correlation, which could indicate that its fast early decline may be dictated by a different physical mechanism than fallback. 
   After performing a volumetric correction to a sample of 30 TDEs observed between 2018 and 2020, we conclude that faint TDEs are not rare by nature;     they should constitute up to $\sim$ 50 -- 60 \% of the entire population and their numbers could alleviate some of the tension between the observed and theoretical TDE rate estimates. We calculate the optical TDE luminosity function and we find a steep power-law relation $dN/dL_{g} \propto {L_{g}}^{-2.36\pm0.16}$.}
%, and a lower-limit for the TDE rate of $\dot{N}\sim2\times 10^{-8}$ Mpc$^{-3}$ yr$^{-1}$.}
% $\approx 3\times10^{-5}\, \rm galaxy^{-1}\, yr^{-1}$.}
   

   \maketitle
%
%________________________________________________________________

\section{Introduction} \label{sec:intro}

The immense gravitational force of supermassive black holes (SMBHs) located in the nuclei of galaxies can lead to the tidal disruption of orbiting stars. When the trajectory of an orbiting star (with stellar radius  R$_{\rm *}$ and   mass  M$_{\rm *}$) intersects the tidal radius R$_{\rm t}$ of the SMBH (with mass  M$_{\rm BH}$), the pericenter distance of the star R$_{\rm p}$ becomes smaller than the tidal radius defined as R$_{\rm t}$~$\approx$~R$_{\rm *}$(M$_{\rm BH}$/M$_{\rm *}$)$^{1/3}$ \citep{Hills1975}. The gravitational field of the SMBH causes a large spread in the specific orbital binding energy of the star, which ends up being much greater than its mean binding energy, leading to the star being ripped apart \citep{Rees1988} in a tidal disruption event (TDE). 

The stellar debris is stretched into a thin elongated stream, half of which escapes the system in unbound orbits, while the rest stay bound and start to fall back toward  the SMBH on highly eccentric orbits \citep{Rees1988,Evans1989}. A strong, transient flare arises \citep{Lacy1982,Rees1988,Evans1989,Phinney1989} as the debris circularizes around the black hole with luminosites of L$_{\rm bol} \sim$ 10$^{41-45}$ erg~s$^{-1}$, which are sometimes   super-Eddington \citep{Strubbe2009,Lodato2011}, that is, their luminosity exceeds the Eddington luminosity (which depends linearly on the BH mass). Theory   predicted TDEs almost five decades ago \citep{Hills1975}; however, they started being observed about  two decades ago, first in the X-ray regime \citep{Komossa1999}, then in the ultraviolet (UV) \citep{Gezari2006}, and finally in the optical wavelengths \citep{Gezari2012}. Moreover, TDEs have been discovered in the mid-infrared \citep{Mattila2018,Kool2020,Jiang2021,Reynolds2022} and   in the gamma-ray, X-ray and radio wavelengths; emission attributed to the launch of relativistic jets and outflows (e.g., \citealt{Zauderer2011,VanVelzen2016,Alexander,Goodwin2022}). However, there is a discrepancy of an order of magnitude between the observed rates of TDEs and the theoretical estimations  (see, e.g., \citealt{Magorrian1999,Wang2004,Stone2016} and review by \citealt{Stone2020}).

In the optical TDEs are considered to be bright with peak absolute magnitudes  on the order of $-20\lesssim M \lesssim-19$ \citep{VanVelzen2020}. However, there have been a few nearby TDEs  ($\sim$ 60-70 Mpc) that are significantly fainter, but also faster than expected. iPTF16fnl \citep{Blagorodnova2017,Brown2018,Onori2019} is the faintest TDE to date ($M_{g} = -17.45$~mag), and was also very fast-decaying TDE. AT~2019qiz \citep{Nicholl2020,Hung2020a} was a fast-decaying TDE with a luminosity intermediate between the
bulk of the population and the faint and fast event iPTF16fnl. These two TDEs belong to the spectroscopic class of H+He TDEs with strong \ion{N}{III} lines as well \citep{Leloudas2019,VanVelzen2020,Charalampopoulos2022}. Two more TDEs of similar brightness to AT~2019qiz are AT~2018ahl \citep{Hinkle2022} and AT~2020neh (\citealt{Angus2022}; this study was published when our work was in a very late stage). 
The latter features   the fastest rise time among all TDEs, and it has been suggested that it might originate from an intermediate-mass  black hole.

Although TDEs are considered to have a steep luminosity function ($dN/dL_{g} \propto {L_{g}}^{-2.5}$, \citealt{Velzen2018}), we are biased in discovering bright ones in flux limited surveys. This is an effect of the Malmquist bias, which describes the preferential detection of intrinsically brighter objects in brightness-limited surveys \citep{Malmquist1922}. Recently, a photometric sample study \citep{Hammerstein2022} presented a uniformly selected sample of 30 spectroscopically classified TDEs observed by the Zwicky Transient Facility Phase-I survey operations ($\sim$ 2.5 years) and the mean peak absolute magnitude of the sample is ${<M_{g,\,\rm peak}>=-19.85}$~mag. A faint and fast TDE in this sample, AT~2020wey, was first discovered by the Zwicky Transient Facility \citep[ZTF;][]{Bellm2019,Masci2019,Patterson2019} as ZTF20acitpfz on 2020 October 13 \citep{Nordin2020} at coordinates R.A. = 09:05:25.880, Dec. = +61:48:09.18. It is coincident with the center of the galaxy \mbox{SDSS J090525.86+614809.1} (also known as \mbox{WISEA J090525.88+614809.2}), an $r=16.71$~mag galaxy found in the Sloan Digital Sky Survey (SDSS)  catalog \citep{Gunn2006}. The transient was also independently reported by the Asteroid Terrestrial impact Last Alert System \mbox{\citep[ATLAS;][]{Tonry2018}} as ATLAS20belb on 2020 October 23 UT, and by Gaia Science Alerts \citep{Hodgkin2013} as Gaia20fck on 2020 November 06 UT. The transient appeared in our custom query in \texttt{Lasair} alert broker\footnote{\url{https://lasair.roe.ac.uk/}} \citep{Smith2019}, a query created to look for nuclear transients within a catalog of quiescent Balmer-strong (QBS) galaxies, otherwise known as E+A galaxies  (i.e., an elliptical galaxy showing absorption lines from A-type star atmospheres). TDEs are over-represented in such galaxies by a factor of 30–35 \citep{Arcavi2014,French2016,Graur2018,French2020}. This catalog of E+A galaxies was retrieved from \citet{French2018} and was implemented in the query. The transient was subsequently classified as a TDE by \citet{Arcavi20wey} on 2020 October 23 UT, based on a spectrum taken with the FLOYDS-N instrument mounted on the Faulkes Telescope North. The spectroscopic redshift of AT~2020wey, measured from narrow galactic absorption lines in the TDE spectrum, is ${z = 0.02738}$. This corresponds to a distance of 124.3 Mpc assuming a $\Lambda$CDM cosmology with \mbox{H$_{0}$ = 67.4\,km\,s$^{-1}$ Mpc$^{-1}$}, ${\Omega_{\rm m}}$ = 0.315, and ${\Omega_{\rm \Lambda}}$ = 0.685 \citep{Aghanim2020}.

\indent In this paper we present the follow-up and thorough photometric and spectroscopic analysis of the faint and fast TDE candidate AT~2020wey.
We describe our observations and data reduction in Sect. \ref{sec:observations}. We present a comprehensive analysis of the photometric and spectroscopic properties of AT~2020wey in Sect. \ref{sec:analysis}, and we discuss their implications in Sect. \ref{sec:discussion}. In Sect. \ref{sec:conclusion} we present our summary and conclusions.

\section{Observations and data reduction} \label{sec:observations}

Throughout this work we assume a \citet{Cardelli1989} extinction law with $R_{V}$ = 3.1 and use a foreground Galactic extinction of $A_{V}$ = 0.127 mag \citep{Schlafly2010} to de-redden our photometry and spectra.

\subsection{Ground-based imaging} \label{subsec:gbi}
Well-sampled host subtracted light curves of AT~20202wey were obtained by the ZTF public survey in the $g$ and $r$ bands, which were accessed using the \texttt{Lasair} alert broker. Las Cumbres Observatory (LCO) $BVgri$-band data were obtained using the Sinistro cameras on Las Cumbres 1m telescopes. PSF fitting was performed on the    host-subtracted  images using the \texttt{lcogtsnpipe} pipeline \citep{Valenti2016}, which uses \texttt{HOTPANTS} \citep{Becker2015} for the subtraction, with template images also obtained  at Las Cumbres after the event became non-detectable. $BV$-band photometry was calibrated to the Vega system using APASS, while $gri$-band photometry was calibrated to the AB system using the Sloan Digital Sky Survey \citep{Smith2002}. We measured the TDE position in a LCO V-band image taken $\sim$ 5 days before the peak of the optical light curve of the transient (MJD=59146.4) and the centroid of the galaxy taken as a template image in the same filter, about  six months later. We measure an offset of $0.096''$ in R.A. and $0.111''$ in Dec. leading to an offset of $0.147''$ from the galactic nucleus. This corresponds to an offset of $\sim$ 80 parsecs at the redshift of the transient (z=0.02738), making AT~2020wey a nuclear transient (with a root mean square error of $\sim$ $0.18''$ in the images, i.e., $\sim$ $\pm$100 parsecs).

Finally, we obtained two epochs of Campo Imperatore photometry and the $g$ and $r$ filters. The Campo Imperatore Station of the INAF-Astronomical Observatory of Abruzzo has a 90 cm Schmidt telescope, equipped with an Apogee 4096$\times$4096 CCD, covering a field of view of 1.15$\times$1.15 square degrees and a $ugriz$ Sloan filter set. For our analysis and plots, all the photometric data were converted to the AB system.

\subsection{Swift UVOT photometry} \label{subsec:uvot_phot}

Target-of-opportunity observations spanning 39 epochs (PIs Arcavi and Charalampopoulos) were obtained with the UV-Optical Telescope (UVOT) and
X-ray Telescope (XRT) on board the Neil Gehrels Swift Observatory
(Swift). The UVOT data were reduced using the standard pipeline available in the \texttt{HEAsoft} software package.\footnote{\url{https://heasarc.gsfc.nasa.gov/docs/software/heasoft/}} Observation of every epoch was conducted using one or several orbits. To improve
the signal-to-noise ratio (S/N) of the observation in a given band in a particular epoch, we co-added all orbit data for that corresponding epoch using the \texttt{HEAsoft} routine \texttt{uvotimsum}. We used the routine \texttt{uvotsource}
to measure the apparent magnitude of the transient (using the most recent UVOT photometric zero-points of \citealt{Poole2007}) by performing aperture photometry using a 5 arcsec aperture for the source and a 25 arcsec aperture for the background. 
%This is approximately twice the UVOT point-spread function, ensuring the measured magnitudes capture most of the transient flux while minimizing the host contribution. 
No host galaxy images in the UV are available for subtraction hence we estimated the host contribution using a spectral energy distribution (SED) fit to archival data for this galaxy (details in Sect. \ref{subsec:host_sed}). The computed synthetic UVOT host magnitudes can be found in Table \ref{tab:synth_phot}.

\begin{table}
\centering
 \caption{Synthetic host-galaxy magnitudes}
 \label{tab:synth_phot}
 \begin{tabular}{ccc}
  \hline
  Filter & Magnitude & Magnitude Uncertainty\\
  \hline
  UVOT V & 16.811 & 0.028 \\
  UVOT B & 17.547 & 0.054 \\
  UVOT U & 18.862 & 0.063 \\
  UVOT W1 & 20.370 & 0.104 \\
  UVOT M2 & 21.216 & 0.211 \\
  UVOT W2 & 21.368 & 0.249 \\
  \hline
 \end{tabular}\\
\begin{flushleft} Synthetic UVOT magnitudes of the host galaxy of AT~2020wey \mbox{SDSS J090525.86+614809.1} derived with \texttt{PROSPECTOR}. All magnitudes are presented in the AB system. \end{flushleft}
\end{table}

\begin{figure*}
\centering
\includegraphics[width=1 \textwidth]{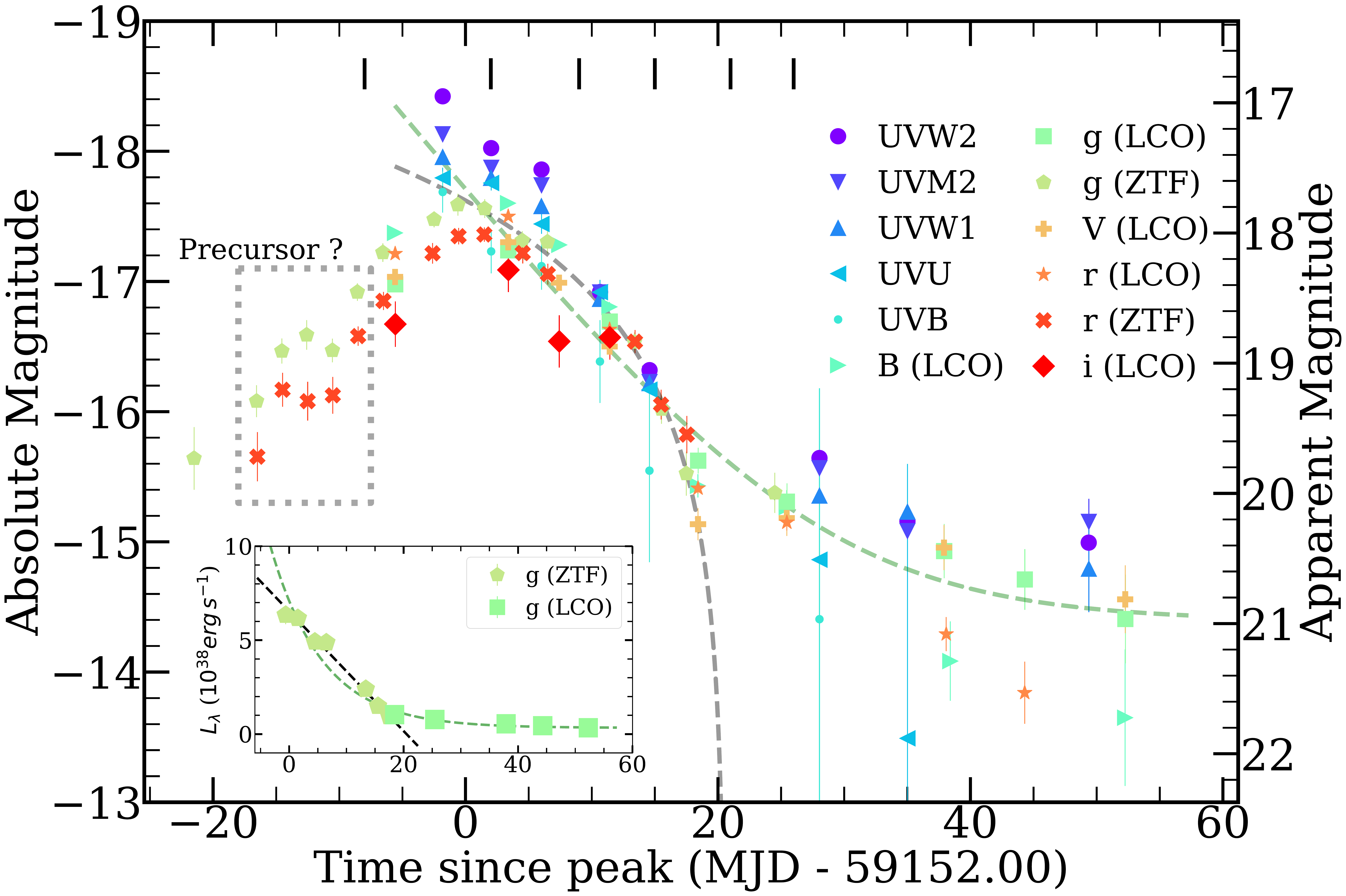}
\caption{ De-reddened and host-subtracted  light curves of AT~2020wey. After the peak, it seems that the light curve goes under two distinct phases: one until $\sim$ 20 days and another after 20 days. The inset shows fits in the $g$ band (dashed lines), to the first phase (from peak to $\sim$ 20 days), and to the entire evolution. The first phase is fit with a linear fit (black line) and the full decline is fit with an exponential (green line). The same fits are plotted in the main plot (dashed lines) but in magnitude space. The exponential fit describes the late-time data nicely, but fails in following the early phase decline. As the event rises to peak, there seems to be a pre-peak precursor (ZTF $g$ and $r$ bands), never observed before in a TDE, between $\sim$ $-18$ days and $-10$ days before the peak.}
\label{fig:photometry}
\end{figure*}

The complete, host-subtracted and de-reddened UV and optical light curves from Swift, LCO, ZTF, and Campo imperatore are shown in Fig. \ref{fig:photometry} and all magnitudes are listed in Tables \ref{tab:ZTF_phot}, \ref{tab:LCO_phot}, and \ref{tab:SWIFT_phot} in the Appendix.

\subsection{X-ray and infrared upper limits} \label{subsec:uplims}

We looked for X-ray emission from the position of the source in the XRT data, and we report that AT~2020wey was not bright in X-rays. After stacking all available observations (eight epochs) we place an upper limit at $8\times10^{-4}$ counts per second. Using the webPIMMS tool\footnote{\url{ https://heasarc.gsfc.nasa.gov/cgibin/Tools/w3pimms/w3pimms.pl}} and assuming a Galactic column density $N_{H}$ of $6.22\times10^{20}$~cm$^{-2}$, we convert the count rate to an unabsorbed flux upper limit of $3.91\times 10^{-14}\,\rm erg\,cm^{-2}\,s^{-1}$ (0.3–10~keV, assuming a blackbody model with kT=0.1 keV), which translates to an upper limit for the X-ray luminosity of $\sim 6.64\times 10^{40}\,\rm erg\, s^{-1}$ (a power-law model with $n=2$ results in an upper limit of $\sim 6.09\times10^{40}\,\rm erg\, s^{-1}$).  %We have also looked at Neutron star Interior Composition Explorer (NICER) data and we did not detect any X-ray activity there either. 

We also looked for mid-infrared data from the \mbox{NEOWISE} survey \citep{Mainzer2011,Mainzer2014} by the Wide-field Infrared Survey Explorer WISE satellite in its W1 and W2 channels. We do not detect any transient activity up to approximately one year after the event and we measure the host galaxy levels at $16.85 \pm 0.02$~mag for W1, and at $17.43 \pm 0.04$ mag for W2 (AB system). That places three-sigma upper limits on the IR detection of the transient at $16.67$~mag for W1 and $17.02$~mag for W2.

\begin{table*} 
\renewcommand{\arraystretch}{1.2}
\setlength\tabcolsep{0.32cm}
\fontsize{10}{11}\selectfont
\begin{center}
\caption{Spectroscopic observations of AT~2020wey}\label{tab:spec_log}
\begin{tabular}{ccrccccc}
\hline
      UT date & 
    MJD & 
    Phase$^{a}$ & 
    Telescope+Instrument &
    Grism/Grating &
        Slit Width &
        Airmass &
    Exposure Time  \\
    (yy-mm-dd)   &  (days)    &   (days)  &     &   &   (arcsec)  &  (arcsec)   &  (s) \\

2020-10-22      &   59144 & $-$7.8 & LCO+FLOYDS         & red/blu                       &  2      & 1.8 & 3600     \\
2020-11-01      &   59154 & 1.9 & LCO+FLOYDS    & red/blu                       &  2      & 1.5 & 3600     \\
2020-11-08      &   59161 &  8.8  & NOT+ALFOSC  & GR\#4                         &  1      & 1.4 &  1800 \\
2020-11-14      &   59167 & 14.6 & LCO+FLOYDS   & red/blu                       &  2      & 1.4 &  3600     \\
2020-11-20      &   59173 &  20.4  & NOT+ALFOSC         & GR\#4                         &  1      & 1.2 &  1350 \\
2020-11-25      &   59178 & 25.3 & LCO+FLOYDS   & red/blu                       &  2      & 1.5 &  3600     \\
2022-02-22      &   59632 &  463.3  & NOT+ALFOSC        & GR\#4                         &  1      & 1.3 &  2700 \\
\hline
\end{tabular}
\\[-10pt]
\end{center}
$^{a}$With respect to the date of $r$-band maximum ($\mathrm{MJD} = 59152$) and given in the rest frame of AT~2020wey ($z=0.02738$).
\end{table*}

\begin{figure*}
        \centering
        \begin{subfigure}[b]{1\textwidth}
            \centering
        \includegraphics[width=0.492 \textwidth]{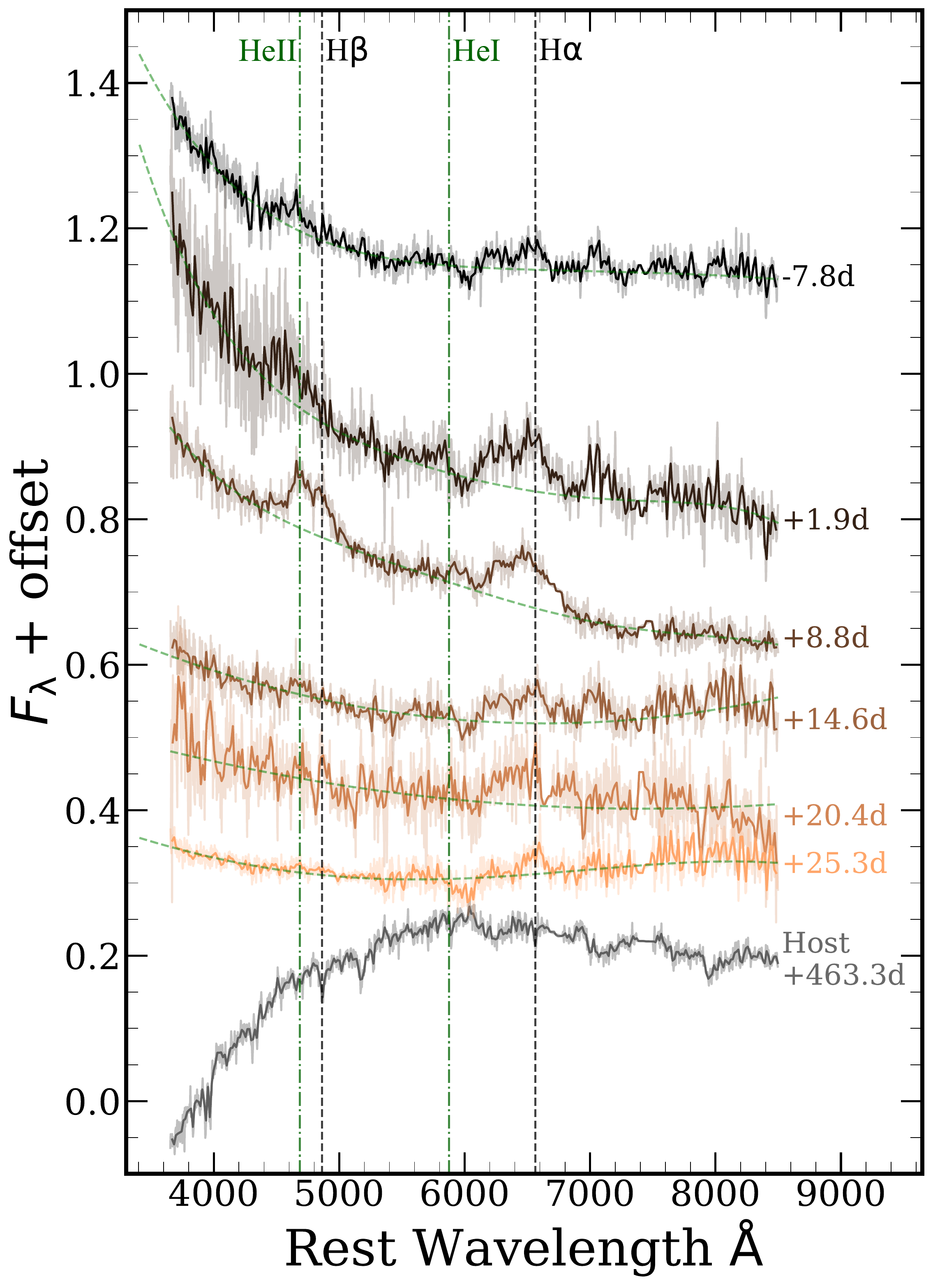}
        \includegraphics[width=0.492 \textwidth]{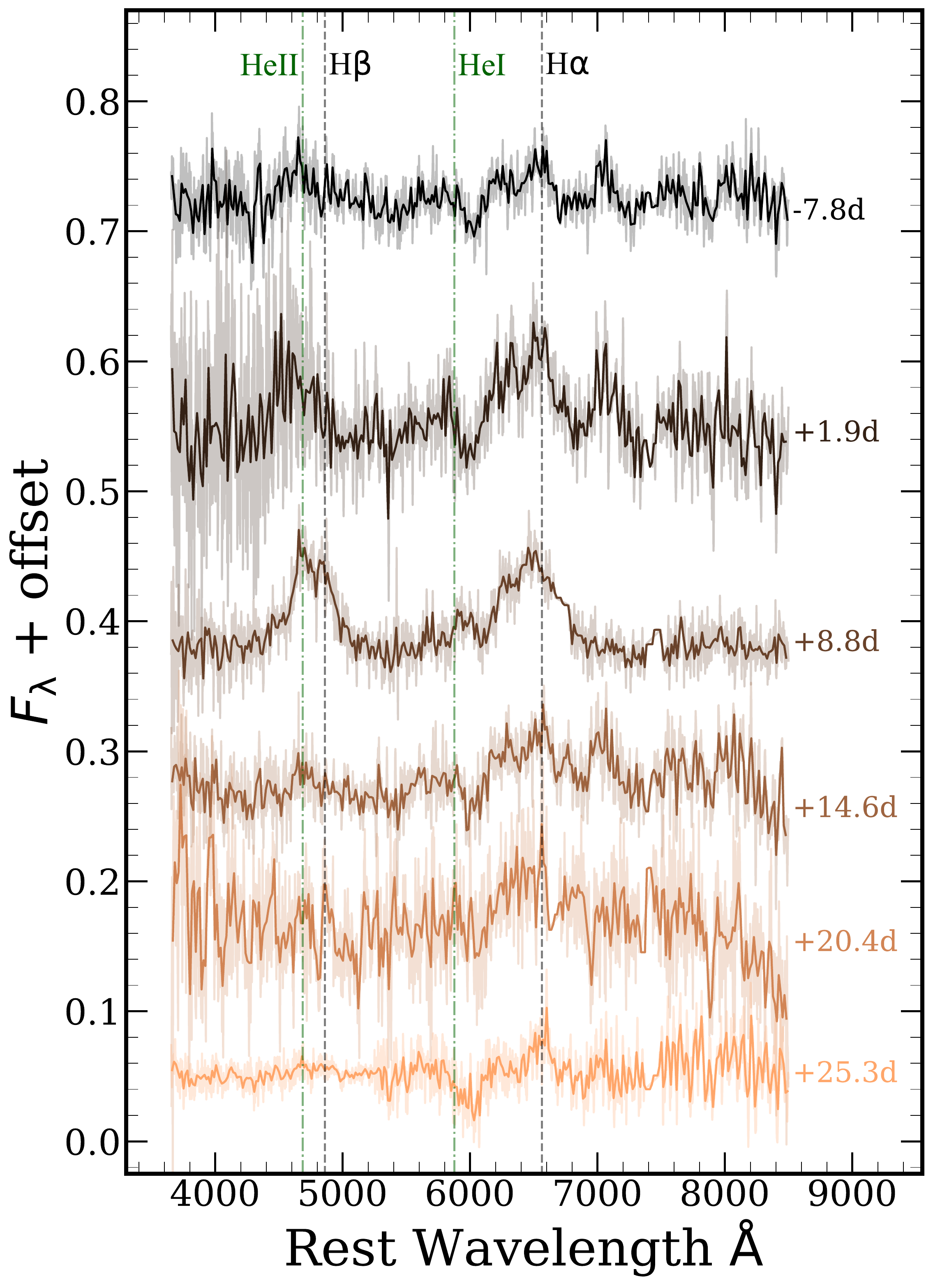}
            % \caption[Network2]%
            % {{\small Network 1}}    
            % \label{fig:off_sub}
        \end{subfigure}
        % \vskip\baselineskip
        \caption{Spectral evolution of AT~2020wey. Left panel:  De-reddened and host-subtracted  spectra of AT~2020wey. The spectrum at $+463.3$ days is the host galaxy spectrum used for host subtraction. The green dashed lines are the polynomial fits to the continuum. Right panel: De-reddened, host-subtracted,  and continuum-subtracted spectra of AT~2020wey.}
        \label{fig:atlas}
    \end{figure*}

\subsection{Optical spectroscopy} \label{subsec:opt_spec}

We   obtained low resolution spectra of AT~2020wey using the FLOYDS instrument on the LCO\footnote{\url{http://lco.global}} 2m telescope in Siding Spring, Australia \citep{Brown2013} and the Alhambra Faint Object Spectrograph and Camera (ALFOSC) mounted on the 2.56m Nordic Optical Telescope (NOT) located at La Palma, Spain. The LCO spectra were extracted, and flux and wavelength calibrated using the \texttt{floyds\_pipeline}\footnote{\url{https://github.com/LCOGT/floyds_pipeline}} \citep{Valenti2014}. The ALFOSC spectra were reduced using the spectroscopic data reduction pipeline \texttt{PyPeIt} \footnote{\url{https://github.com/pypeit/PypeIt}} \citep{Prochaska2020}. Although an archival SDSS spectrum existed, we obtained a high S/N host galaxy spectrum with the NOT at 463~d after peak (rest frame) in order to host-subtract our spectra with the same instrument. A spectroscopic log can be found in Table \ref{tab:spec_log} and the host-subtracted  spectral series along with the host is shown in the left panel of Fig. \ref{fig:atlas}. We note that there were four more LCO spectra acquired, one on November 08 and three within January. We do not show those spectra in Fig. \ref{fig:atlas} nor include them in the analysis as we have a higher S/N spectrum from the NOT on November 08, and the late ones were acquired after the optical emission had reached host levels. The host subtraction was performed similarly to \citet{Charalampopoulos2022}.  

% \begin{figure}
% \centering
% \includegraphics[width=0.47 \textwidth]{figures/atlas_offset_with_host.pdf}
% \caption{
% De-reddened and host-subtracted  light curves of AT~2020wey.
% }\label{fig:atlas}
% \end{figure}

% \begin{figure}
% \centering
% \includegraphics[width=0.47 \textwidth]{figures/atlas_offset_contsub.pdf}
% \caption{
% De-reddened, host-subtracted  and continuum-subtracted light curves of AT~2020wey.
% }\label{fig:atlas_cont_sub}
% \end{figure}

\section{Analysis} \label{sec:analysis}

\subsection{Host galaxy and SED model fit} \label{subsec:host_sed}

The host galaxy (presented in the left panel of Fig. \ref{fig:atlas}) shows Balmer absorption lines and lacks emission lines like H$\alpha$ and \ion{O}{II}, which is typical of quiescent Balmer-strong galaxies. It was spectroscopically (SDSS archival spectrum) classified by \citet{French2018} as a quiescent Balmer-strong galaxy, with H$\alpha$ equivalent width of 0.08 \AA\, (in emission) and a Lick H$\delta_{A}$ of 2.92 (in absorption).

\begin{figure}
  \centering
  \includegraphics[width=0.47 \textwidth]{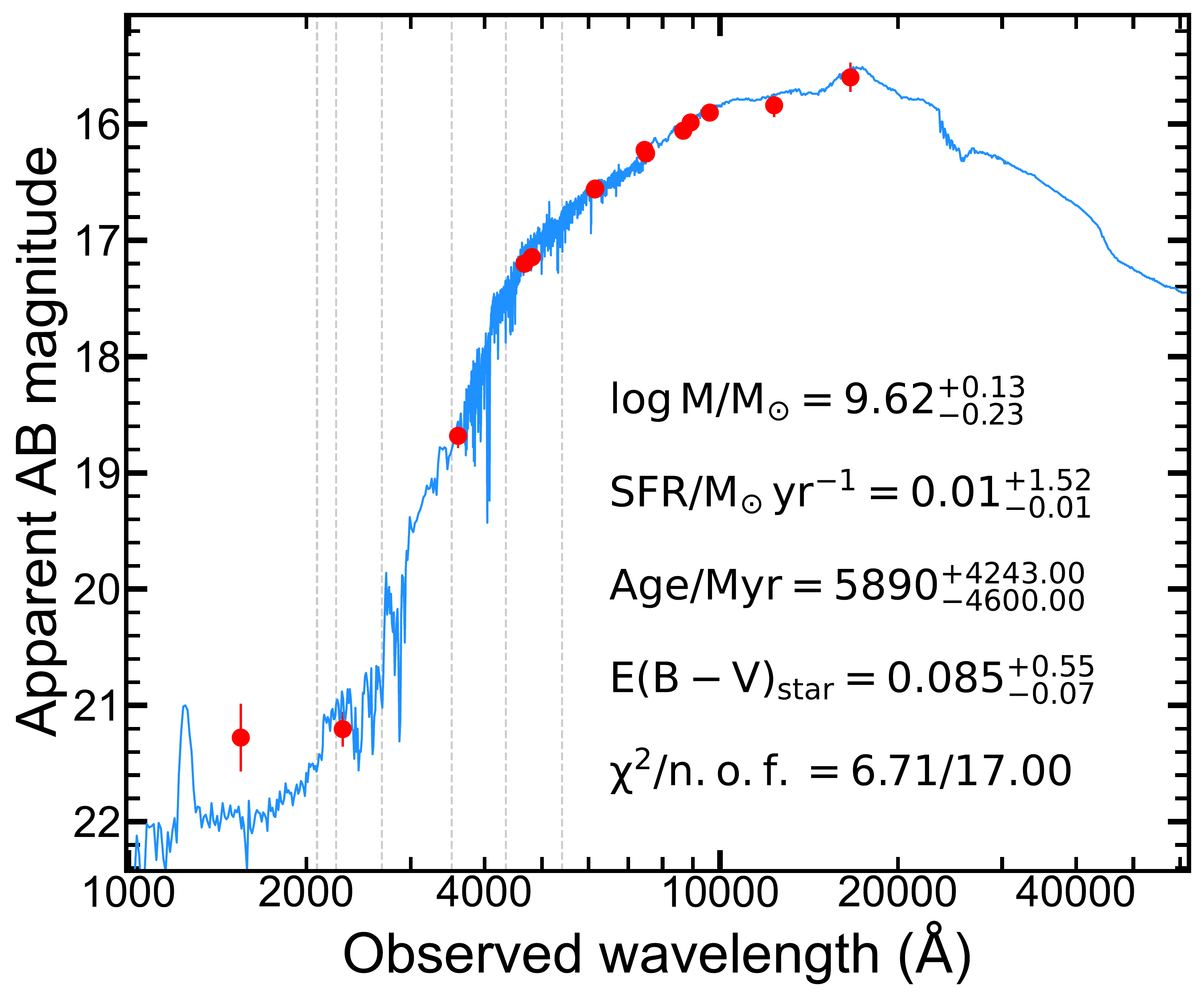}
  \caption{ Host galaxy SED fit with \texttt{PROSPECTOR}. The dashed vertical lines show the effective wavelengths of the Swift UVOT filters for which  synthetic magnitudes have been derived (see Table \ref{tab:synth_phot}).}
  \label{fig:sed_fit}
\end{figure}

We retrieved archival photometry of the host galaxy from the PanSTARRS catalog \citep{Huber2015} in the $g,\,r,\,i,\,z,\,y$ filters and from the Sloan Digital Sky Survey (SDSS; \citealt{York2000}) in the $g,\,r,\,i,\,z$ filters. Furthermore, we retrieved archival photometry from the 2 Micron All Sky Survey (2MASS; \citealt{Skrutskie2006}), and the Galaxy Evolution Explorer (GALEX; \citealt{Martin2005}). We used \texttt{PROSPECTOR} \citep{Leja2017} in order to fit archival photometry and produce the best-fitting SED model (see \citealt{Schulze2020} for a description of the methods), which is shown compared to the archival photometry in Fig. \ref{fig:sed_fit}. The SED model includes a stellar component and emission from the \ion{H}{II} regions. We used the software package \texttt{Lambda Adaptive Multi-Band Deblending Algorithm in R} \citep{Wright2016} to obtain matched-aperture photometry from the far-UV to the near-infrared, and included the entire host galaxy light in order to quantify the properties of the host galaxy. We find a stellar mass of ${\log (M_*/M_\odot) = 9.62^{+0.13}_{-0.23}}$, a star formation  rate of ${\rm SFR=0.01^{+1.52}_{-0.01}\msun yr^{-1}}$ corresponding to a low specific star formation rate of ${\log {\rm sSFR/\msun yr^{-1}}=-11.76^{+1.52}_{-0.01}}$. The reported values and uncertainties are the median and 16th and 84th percentiles of the marginalized posterior distributions. These values confirm that the host is indeed a quiescent galaxy. The model predicts a color excess of ${E(B-V)_{star}=0.085^{+0.55}_{-0.07}}$ mag. The stellar mass reported by \texttt{PROSPECTOR} is the integral of the star formation history, and so includes stars and stellar remnants. Finally, in order to retrieve the UVOT host magnitudes (needed for host subtraction; see Table \ref{tab:synth_phot}) we used an aperture of 5 arcsecs, similar to what was used for the photometry of UVOT.

\subsection{Photometric analysis} \label{subsec:phot_analysis}

The complete, host-subtracted  and de-reddened UV and optical light curves from Swift, LCO, and ZTF are shown in Fig. \ref{fig:photometry}. After fitting a third-order polynomial around the maximum of the ZTF $r$~band,  we conclude that the date of $r$-band maximum is at $\mathrm{MJD} = 59152.00\pm0.45$. From here on we   use this date as the peak of AT~2020wey. %hence where phase=$0\rm d$. 

\subsubsection{Broadband light curve evolution} \label{subsubsec:g-band}

We study the broadband light curve evolution of AT~2020wey in the $g$~band. The time from the last non-detection to the ZTF $g$-band peak is (22.94 $\pm$ 2.03) days. The rise to peak is nicely covered by the ZTF $g$ and $r$~bands. We note that in both bands there seems to be a variation in the smooth rise of the light curve between $\sim$ $-18$~d and $-10$~d, not identified before in TDEs. This \say{precursor} activity can be interpreted either as a bump (third and fourth points in $g$ band and second point in $r$ band) or as a dip (fourth point in $g$ band and second point  in $r$ band), with the dip being more statistically significant as an outlier. The physical processes that drive the light curve rise of TDEs are still under debate, and hence we cannot conclude on the nature or physical origin of the precursor. However, based on the different scenarios of the TDE emission mechanism (and whether it is seen as a bump or a dip), such a feature could be   evidence of a fallback--rapid accretion episode or, if an outflow drives the early pre-peak emission of TDEs \citep{Nicholl2020}, such a feature could be related to an uneven density distribution of the expanding optically thick outflow. Alternatively, it could arise from debris stream-crossing shocks during circularization, followed by the main peak.
%(powered by accretion and/or an outflow).

After the peak, the light curve undergoes two distinct phases, one until $\sim$ $+$20 days and one after. During the first phase the decay of the light curve is surprisingly fast for a TDE and the luminosity decline is best described by a linear fit in luminosity space (see inset of Fig. \ref{fig:photometry}). During the second phase the event drops close to host galaxy levels and the evolution becomes very slow. %Since ZTF stopped observing it during this phase, we use the LCO $g$-band in order to study the decay rate.  
We fit the combined ZTF and LCO  $g$ band with a power-law model and an exponential model, and we find that the exponential model provides a much better fit to the  data with a reduced chi-square statistic of $\chi^{2}_{\nu}=4.41$ (compared to the $\chi^{2}_{\nu}=19.53$ of the power law). Nonetheless, it cannot describe the data during the first decline phase (until $\sim$ $+$20 days) as accurately as the linear fit ($\chi^{2}_{\nu}=0.62$), as shown in the inset of Fig. \ref{fig:photometry}.

\begin{figure}
  \centering
  \includegraphics[width=0.5 \textwidth]{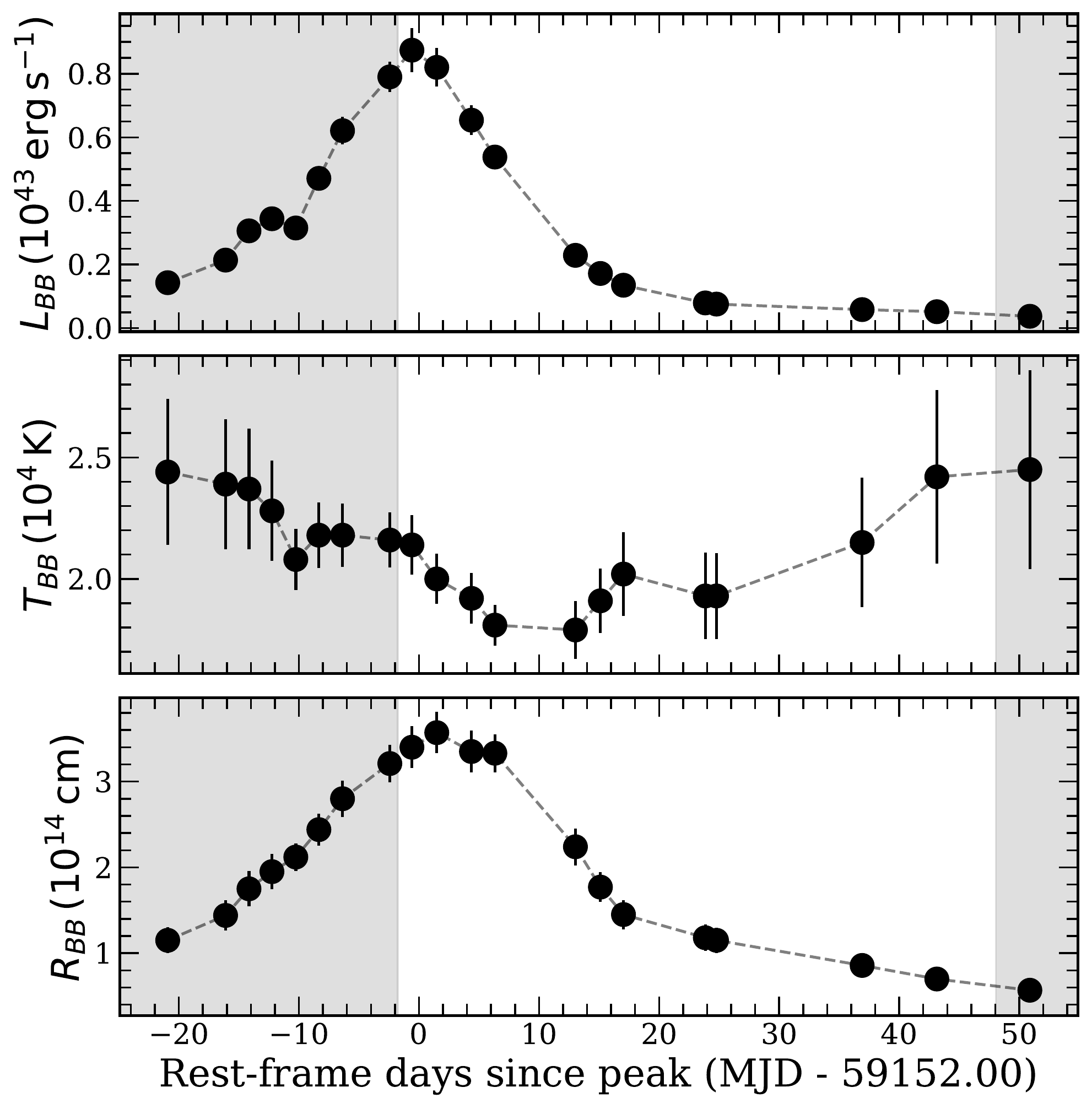}
  \caption{ Bolometric light curve, temperature, and radius evolution derived using \texttt{SUPERBOL}. The shaded regions indicate the time regions where   constant color was assumed for extrapolation purposes.}
  \label{fig:bol}
\end{figure}

\begin{figure}
\centering
\includegraphics[trim={0 0 0 0},clip,width=0.49 \textwidth]{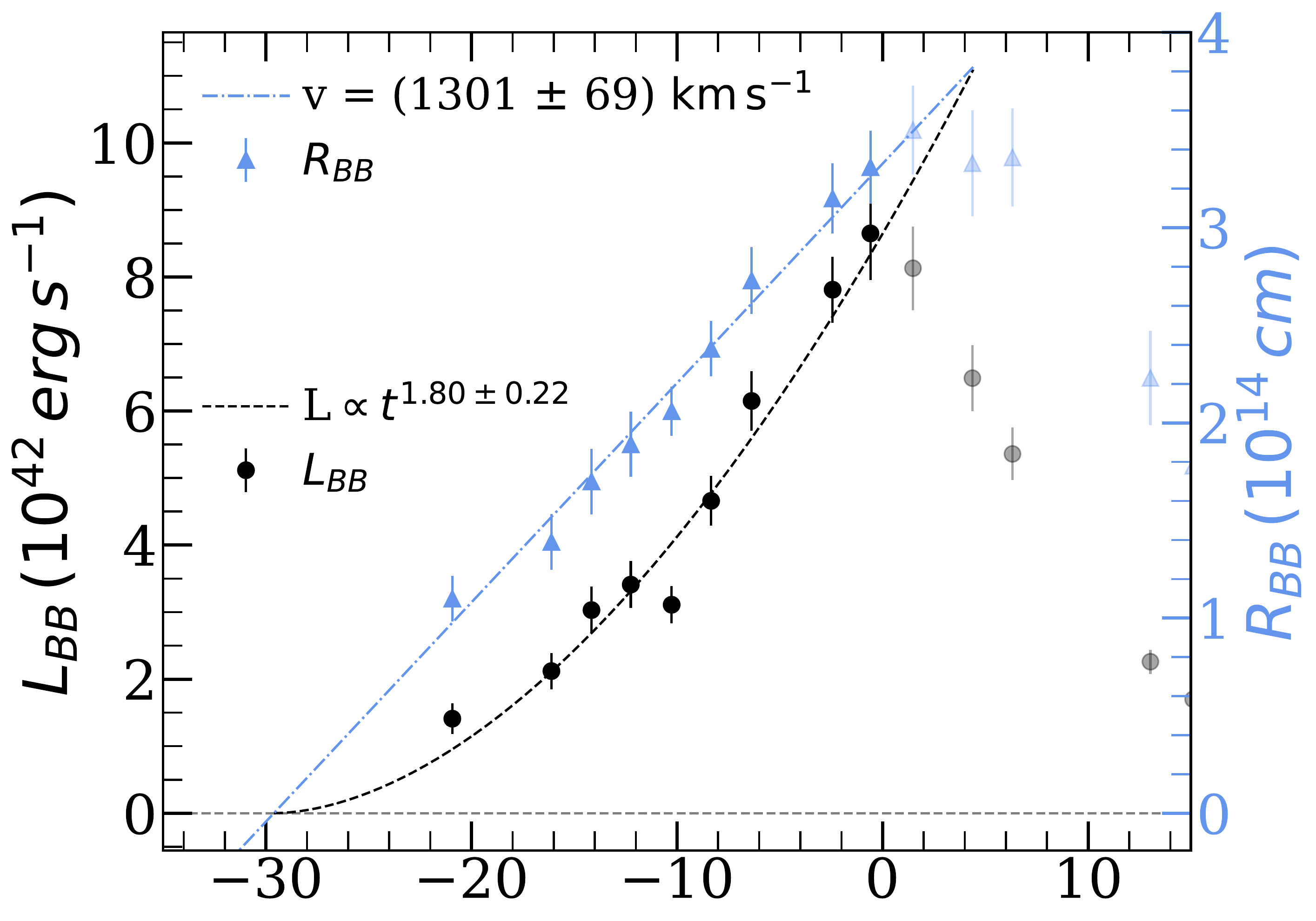}
\includegraphics[trim={0 0 0 0},clip,width=0.5 \textwidth]{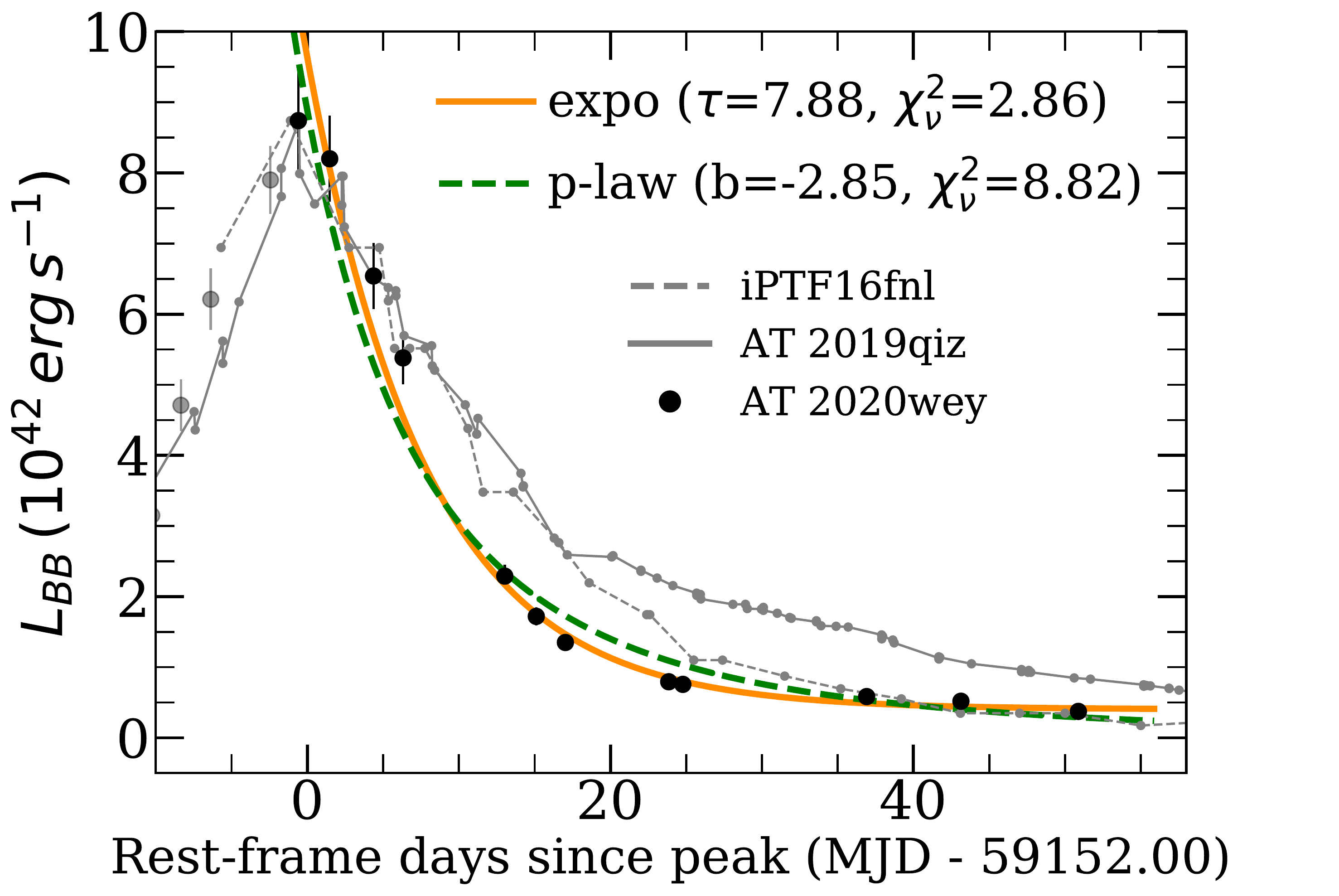}
\caption{Fits to the bolometric light curve during   rise (top) and   decline (bottom) of the light curve. Top: The photospheric radius (blue triangles) before peak grows linearly with a velocity $v\approx1\,300\,\rm km\,s^{-1}$, while the luminosity (black circles) is best fit as $L\propto t^{1.8}$. Bottom: Exponential and power-law fits to the declining bolometric light curve. The decline is much steeper than the canonical $t^{-5/3}$.  For comparison, iPTF16fnl and AT~2019qiz are overplotted  and   their bolometric light curves normalized in order for the three events to have the same peak; AT~2020wey is the fastest declining TDE to date.}
\label{fig:randd}
\end{figure}

\subsubsection{Bolometric light curve} \label{subsubsec:Bol}

We construct the bolometric light curve of AT~2020wey by interpolating the host-subtracted  and de-reddened photometry of each band in time, to any epoch with data in the three UV bands of UVOT (reference filters), using \texttt{SUPERBOL} \citep{Nicholl2018a}. For the interpolation of the light curves we used polynomials of third to fifth order. Then we integrate under the spectral energy distribution (from 1 to 25000 \AA) at each epoch in order to retrieve the luminosity and fit the SED with a blackbody function to estimate the temperature and the radius, and also in order to calculate the missing energy outside of the observed wavelength range. Before the peak, for the epochs where the reference filters were not observed, we extrapolated assuming constant color based on the ZTF $g-r$ color. We find this choice fair, as TDEs have been speculated to show a generally constant color evolution \citep{Holoien2018,VanVelzen2021}. The constant color assumption is further validated by looking at the shape of the strong blue continuum in the spectra, which is very similar in the first three epochs. The bolometric light curve, temperature, and radius evolution are plotted in Fig. \ref{fig:bol}. The epochs for which we used the constant color extrapolation method are shaded in gray.

\begin{figure}
  \centering
  \includegraphics[width=0.5 \textwidth]{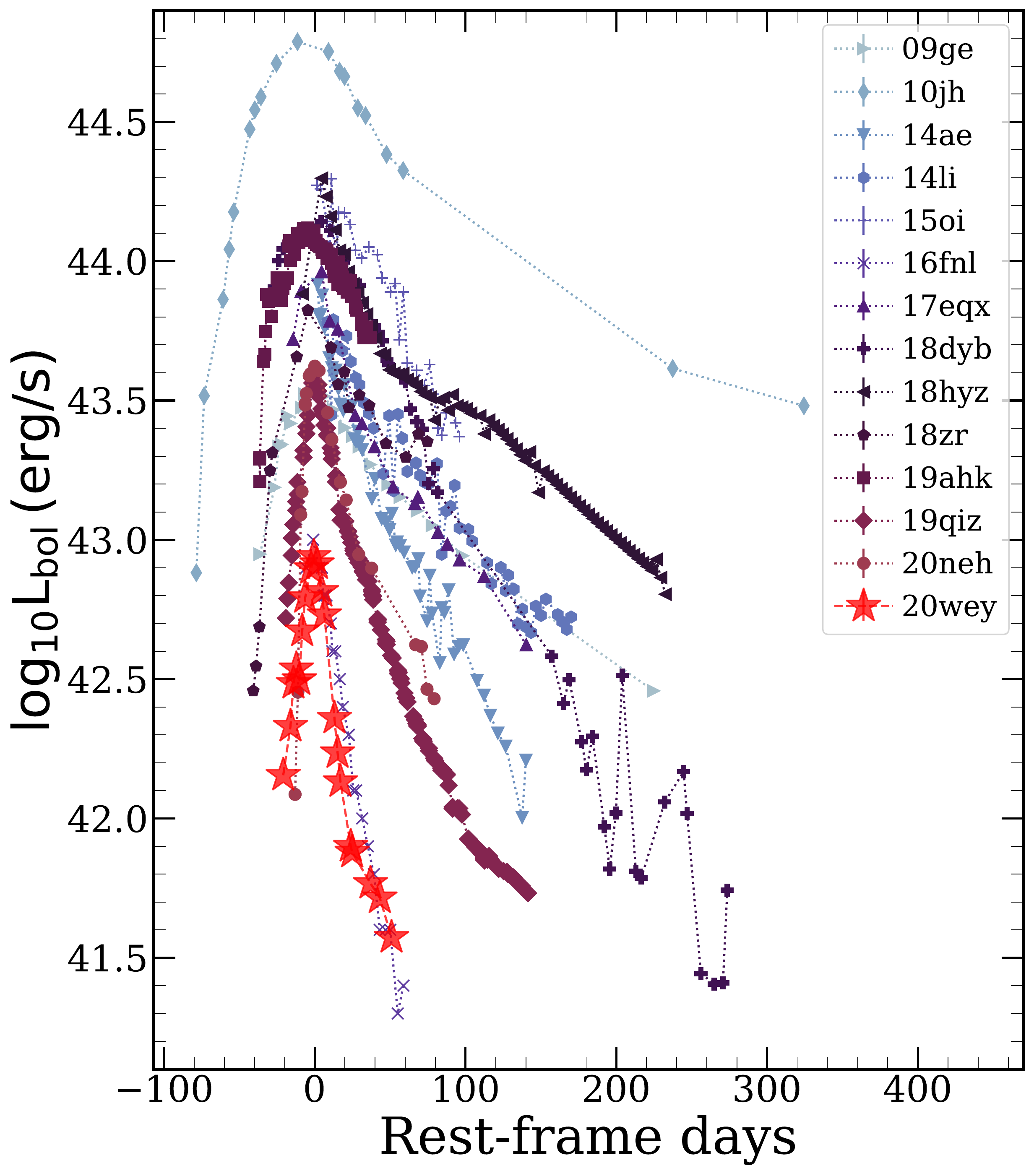}
  \caption{Comparison of the bolometric light curve of AT~2020wey with the TDEs presented in Fig. 10 of \citet{Nicholl2020} and with AT~2020neh \citep{Angus2022}. The resemblance of AT~2020wey with iPTF16fnl is striking. AT~2020neh rises faster, but is brighter and decays more slowly.}
  \label{fig:bol_comp}
\end{figure}

The bolometric light curve analysis shows a blackbody temperature in the range $\sim20\,000\,\rm K-25\,000$~K, that slowly declines  from $\sim25\,000$~K to $\tbb\sim20\,000$~K at peak and is followed by a slow return to $\sim25\,000\rm$~K. The blackbody photosphere expands to the peak value of $\rbb\sim3.5\times10^{14}$~cm before it starts contracting. The expansion of the radius up to the peak is best described by a linear expansion with a constant velocity of $\gtrsim 1\,300\rm\,km\, s^{-1}$ (top panel of Fig. \ref{fig:randd}). \citet{Nicholl2020} finds a linear expansion to the peak radius of the moderately faint TDE AT~2019qiz as well, but with a higher expansion velocity of $2\,200\rm\,km\, s^{-1}$. The higher luminosity TDE AT~2020zso was measured to have an ever higher expansion velocity of $2\,900\rm\,km\, s^{-1}$ \citep{Wevers2022}. Extrapolating back to radius $R_{BB}=0$, we infer the start of the light curve rise to be at 29.6d before peak, similar to the   value that \citet{Nicholl2020} find for AT~2019qiz ($-30.6$d) and   for AT~2020zso ($-35$d \citealt{Wevers2022}). We fit the rise of the bolometric light curve with a power law of the form $L = L_0 [(t-t_0)/\tau]^{\alpha}$ using the \texttt{lmfit}\footnote{\url{https://lmfit.github.io/lmfit-py/}} package \citep{Newville2016}, where a Levenberg--Marquardt algorithm (i.e., least-squares method) was used for fitting. We fix the initial time to $t_{0}=-29.6\rm d$ and we get a best-fit power-law index $a=1.8\pm0.22$, similar to AT~2019qiz ($a=1.99\pm0.01$), AT~2019ahk ($a=2.1\pm0.12$; \citealt{Holoien2019}), and AT~2019azh ($a=1.9\pm0.39$; \citealt{Hinkle2021}). We present this fit alongside the fit to the radius expansion in Fig. \ref{fig:randd}. In the bottom panel of Fig. \ref{fig:randd} we present our fits to the sharply declining part of the bolometric luminosity curve. We fit the decline of the bolometric light curve with a power law and with an exponential decay function ($L \propto e^{-t/\tau}$). We find that both fits can properly describe the data, but the exponential fit returns a better reduced chi-square statistic ($\chi^{2}_{\nu}=2.46$ and $\chi^{2}_{\nu}=7.64$, respectively). The exponential fit has a $\tau=7.86\pm0.48$. \citet{Onori2019} finds that the bolometric luminosity of the faintest TDE to date iPTF16fnl, is also best described by an exponential decay with $\tau=17.6\pm0.2$. That makes AT~2020wey faster and more sharply declining. Our power-law fit returns a power-law index of $b=-2.84\pm0.18$, much steeper than the canonical $L\propto t^{-5/3}$ expected from simple fallback arguments \citep{Rees1988} and even steeper than the fast AT~2019qiz ($b=-2.51\pm0.03$). AT~2020neh, which is the fastest rising TDE, has a moderate decline ($b=-1.44\pm0.19$). The declining rates of the  broadband (see Fig. \ref{fig:photometry}) and the bolometric light curve of AT~2020wey are surprisingly sharp and make it the fastest declining TDE to date. 
%Another intriguing feature of this TDE is that the decline timescale is faster than the rising one, something that is rarely observed in transients.
The peak of the bolometric luminosity ${L_{\rm pk}=(8.74\pm0.69)\times10^{42} \rm\,erg\,s^{-1}}$ is very low for a TDE, making it equally faint to iPTF16fnl, the faintest TDE to date. In Fig. \ref{fig:bol_comp} we plot the bolometric light curve compared to other well observed TDEs studied in \citet{Nicholl2020} (see their Fig. 10). We see that AT~2020wey is the faintest discovered TDE along with iPTF16fnl and that the two TDEs seem identical, almost clones of each other in the bolometric luminosity space.

\begin{figure}
  \centering
  \includegraphics[width=0.5 \textwidth]{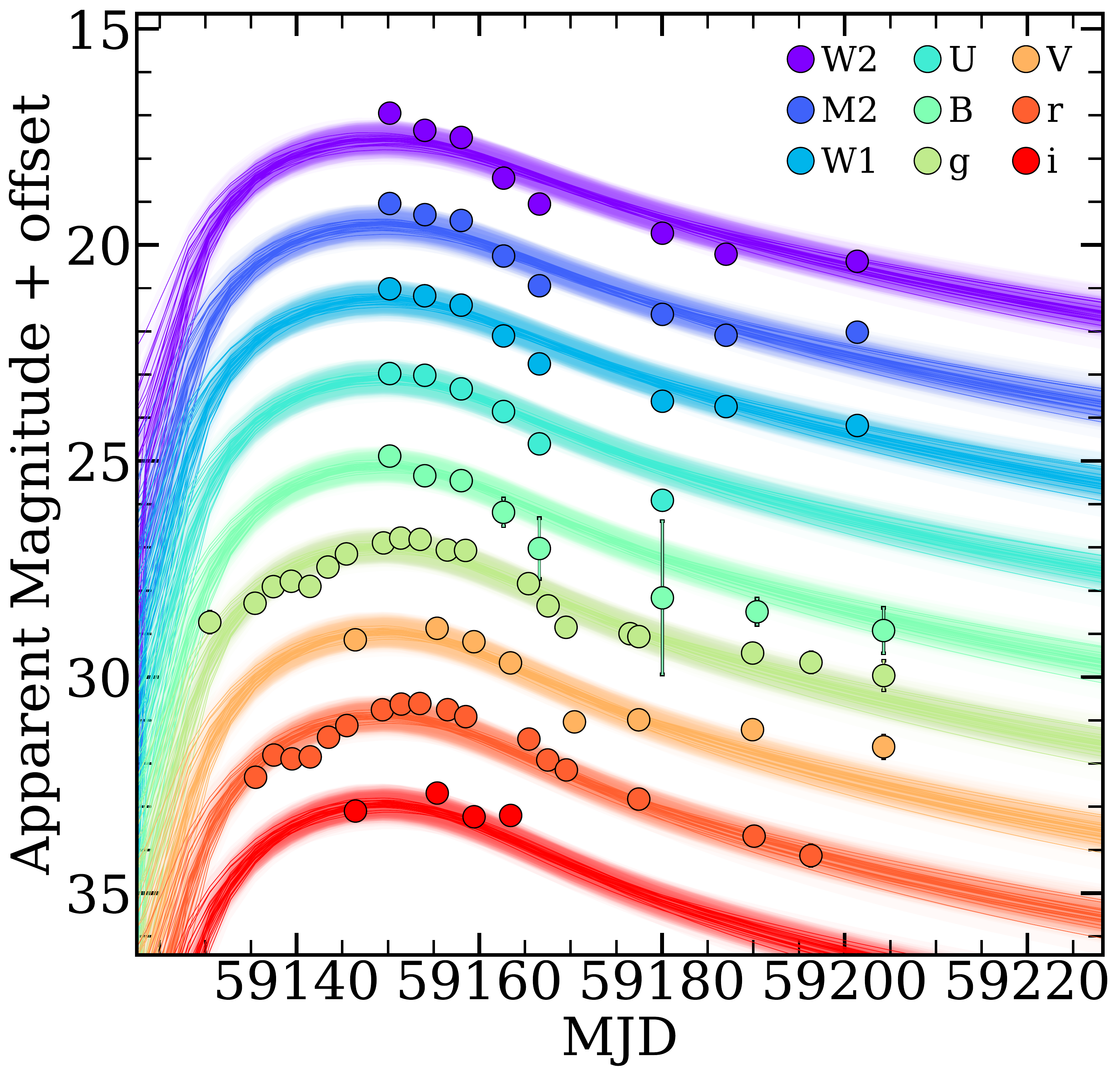}
  \caption{Fits to the multi-color light curve using the TDE model in \texttt{MOSFIT} \citep{Guillochon2018,Mockler2019}.}
  \label{fig:mosfit}
\end{figure}

% Mosfit model
\begin{table}
  \centering
  \begin{tabular}{cccc}
  \hline
  Parameter & Prior & Posterior & Units\\
  \hline
$ \log{(M_\bullet )}$ & $[5, 8]$ & $ 6.46^{+0.09}_{-0.09} $ & M$_\odot$ \\
$ M_*$ & $[0.01, 100]$ & $ 0.11 ^{+0.05}_{-0.02} $  & M$_\odot$ \\
$ b$ & $[0, 2]$ & $ 1.03^{+0.05}_{-0.08}$ &   \\
$ \log(\epsilon) $ & $[-2.3, -0.4] $ & $ -1.95 ^{+0.28}_{-0.2}$ &   \\
$ \log{(R_{\rm ph,0} )} $ & $[-4, 4] $ & $ 1.36 ^{+0.33}_{-0.32} $  &   \\
$ l_{\rm ph}$ & $[0, 4]$ & $ 2.58 ^{+0.29}_{-0.22} $  &   \\
$ \log{(T_v )} $ & $[-3, 3] $ & $ -1.1^{+1.06}_{-1.28} $ & days  \\
$ t_0 $ & $[-150, 0]$ & $  -11.78 ^{+1.87}_{-2.13}$ & days  \\
$ \log{(n_{\rm H,host})}$ & $[19, 23]$ & $ 20.95 ^{+0.05}_{-0.06}$ & cm$^{-2}$ \\
$ \log{\sigma} $ & $[-4, 2] $ & $ -0.46 \pm ^{+0.04}_{-0.04} $  &   \\
  \hline
\end{tabular}
  \caption{Priors and marginalized posteriors for the \texttt{MOSFIT} TDE model. The posterior results are the median of each distribution, and the uncertainties are the 16th and 84th percentiles. These error estimations are purely statistical and further estimates of the systematic uncertainty are provided in \citet{Mockler2019}.}
  \label{tab:mosfit}
\end{table}

\subsubsection{TDE model fit} \label{subsubsec:mosfit}

To estimate the physical parameters of the disruption, we fit our host-subtracted  multi-band light curves using the Modular Open Source Fitter for Transients \citep[\texttt{MOSFIT};][]{Guillochon2018} with the TDE model from \citet{Mockler2019}. This model assumes a mass fallback rate derived from simulated disruptions of polytropic stars by a SMBH of $10^6\,\mathrm{M}_{\odot}$ \citep{Guillochon2014}, and uses scaling relations and interpolations for a range of black hole masses, star masses, and encounter parameters. It generates a bolometric light curve and, in turn, multi-band light curves, which are fitted to the observed data. Finally, it reaches the combination of parameters that yield the highest likelihood match, using one of the various sampling methods. Here, we ran \texttt{MOSFIT} using dynamic nested sampling with \texttt{DYNESTY} \citep{Speagle2020} in order to evaluate the posterior distributions of the model. We list the free parameters of the model (as defined by \citealt{Mockler2019}), their priors, and their posterior probability distributions in Table \ref{tab:mosfit}, with two-dimensional posteriors plotted in Fig. \ref{fig:corner} of the Appendix.

\begin{figure*}
\centering
\includegraphics[width=1 \textwidth]{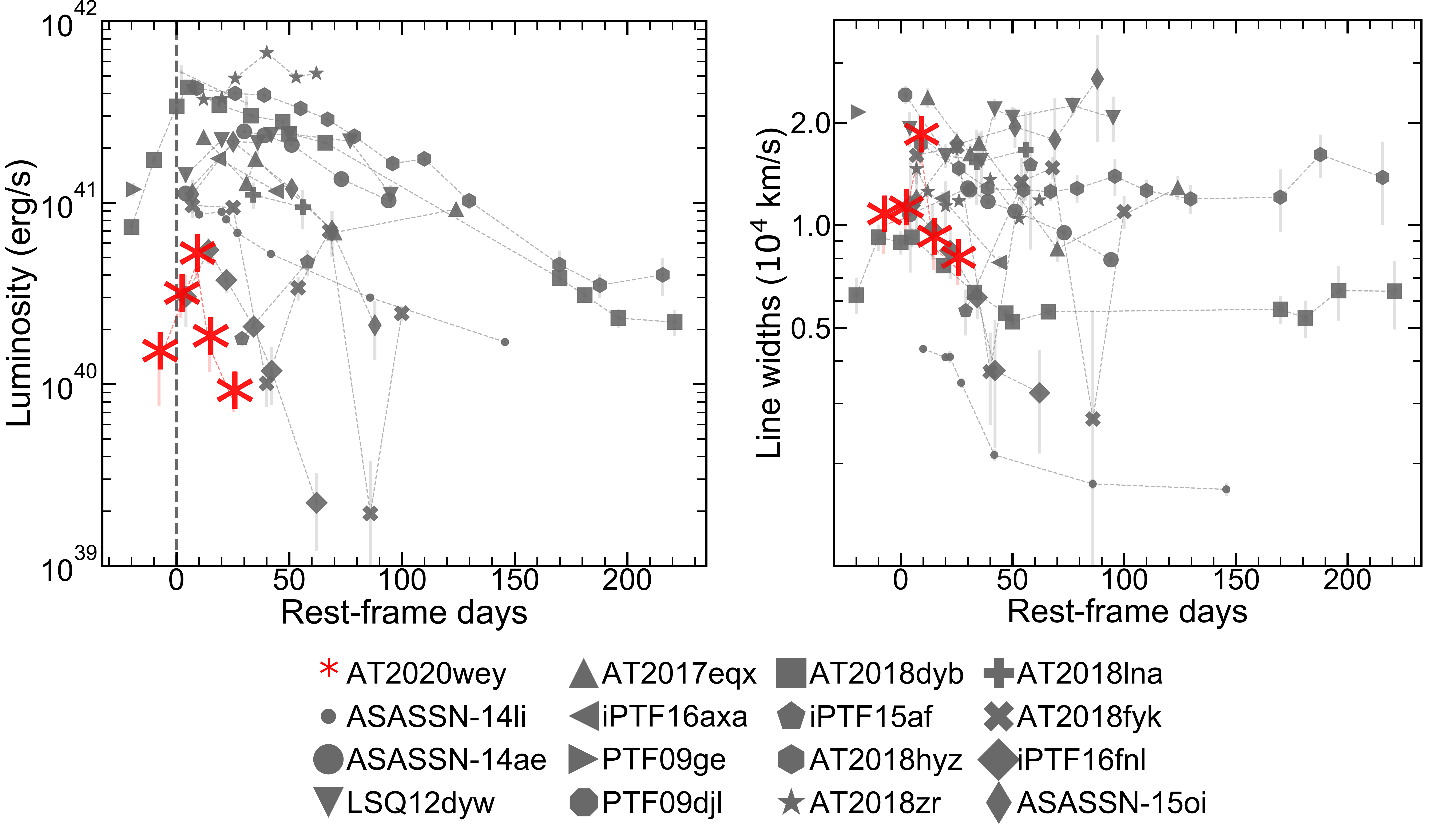}
\caption{H$\alpha$ properties as a function of time. Left panel:   H$\alpha$ luminosity evolution of AT~2020wey as a function of time, compared with the sample of \citet{Charalampopoulos2022}. The dashed vertical line denotes the time of peak or discovery of each TDE. Right panel:   H$\alpha$ line width evolution of AT~2020wey as a function of time, compared with the sample of \citet{Charalampopoulos2022}.}
\label{fig:ha_vs_sample}
\end{figure*}

In Fig. \ref{fig:mosfit} we present the fits to the multi-band light curves of AT~2020wey. The model struggles to reproduce the UV data around the peak and the optical data after +20d post-peak. The latter might be reasonable since this is when the transient has almost faded and reached host levels. For the SMBH and the stellar mass, we find a $ \log{(\mh )}=6.46^{+0.09}_{-0.09} $ and $ M_*=0.11 ^{+0.05}_{-0.02}\msun $, respectively. We note that even though the fit struggles to reproduce the data, the posteriors of these values are narrow. This means that the true uncertainties in our estimates of these quantities must be driven by the systematic errors of the \texttt{MOSFIT} fits. This is addressed by \citet{Mockler2019} who calculated the systematic uncertainties of \texttt{MOSFIT} TDE fitting and found a $\pm$0.2 dex uncertainty in the BH mass and a $\pm$0.66 dex in the stellar mass. Furthermore we find a scaled impact parameter, $b = 1.03^{0.05}_{0.08}$. This scaled impact parameter is defined such that $b\geq1$ leads to a complete disruption and $b=0$ to no disruption (for further discussion see \citealt{Mockler2019}). To summarize all of the above, our \texttt{MOSFIT} posteriors suggest the complete disruption of a sub-solar mass star from a moderate mass supermassive black hole.

\begin{figure}
  \centering
  \includegraphics[width=0.5 \textwidth]{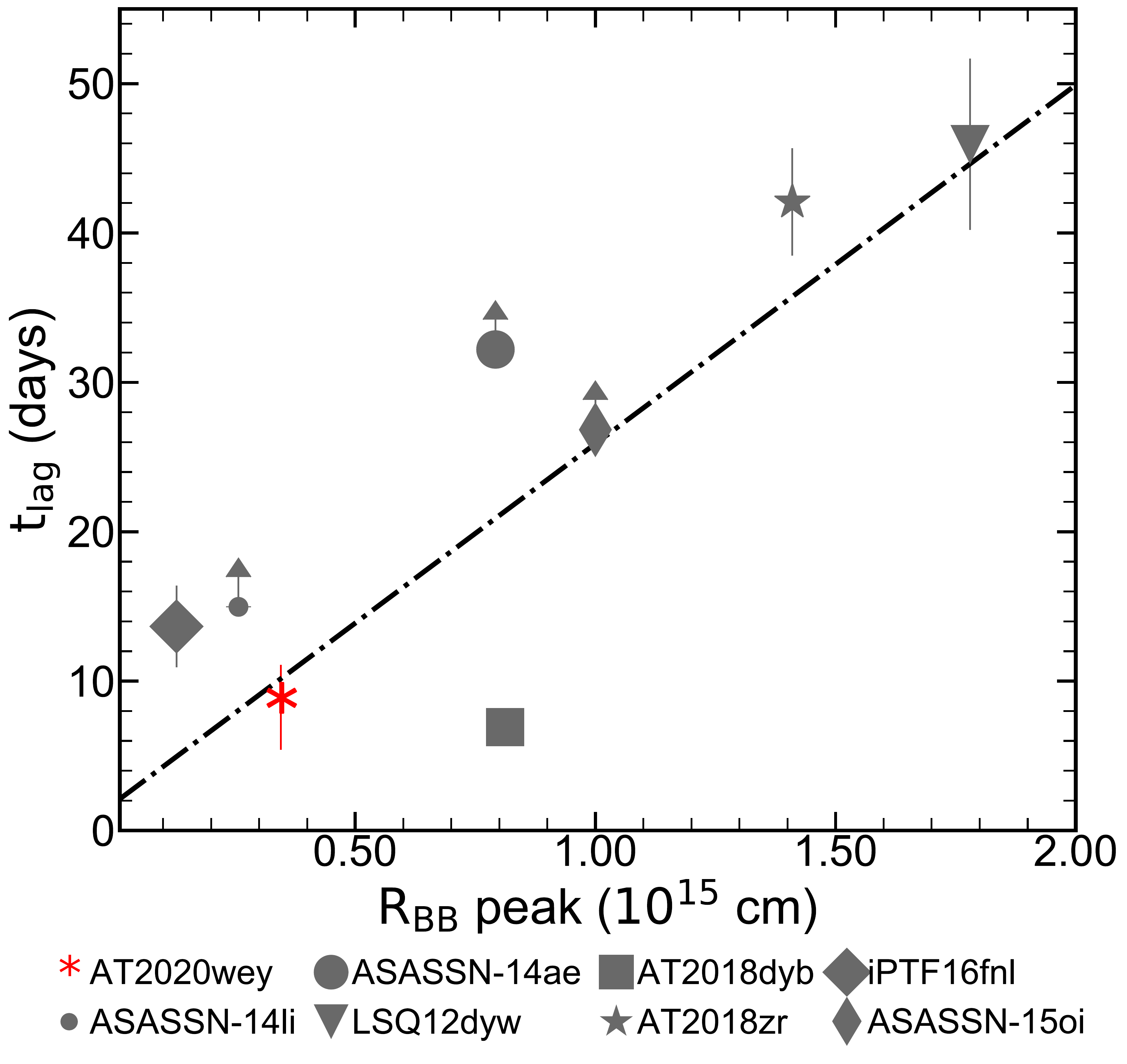}
  \caption{Comparison of the H$\alpha$ lag as a function of the peak of the blackbody radius of AT~2020wey   with the other TDEs \citep{Charalampopoulos2022}. The dot-dashed line is a linear regression fit that does not include the three lower limits.}
  \label{fig:Ha_lag_vs_sample}
\end{figure}

\subsection{Spectroscopic analysis} \label{subsec:spec_analysis}

In the spectra of AT~2020wey, we identify Balmer H$\alpha$ and H$\beta$ lines as well as the typical TDE line \ion{He}{II} 4686\AA\,. A faint \ion{He}{I} 5876\AA\, also appears at 15~d post-peak.  Frequently, the broad \ion{He}{II} 4686 \AA\, line in TDEs is found to be blended with \ion{N}{III} 4640 \AA\ and, depending on the S/N of the spectrum and the width of the lines, it may be impossible to distinguish between the two peaks. The reason why other TDEs (see, e.g., \citealt{Nicholl2020} and \citealt{Leloudas2019}) are classified as \ion{N}{III} rich TDEs (Bowen) is that they show a clear strong peak of \ion{N}{III} $\sim$ 4100\AA,\, which is clearly absent for AT~2020wey in all of our spectra. In order to further quantify the non-detection of the \ion{N}{III} $\sim$ 4100\AA\, line, we measure a two-sigma upper limit of $3.47\times10^{39}$ erg~s$^{-1}$, by integrating the error spectrum of our +8.8-day NOT spectrum over a region of two times the full width at half maximum of the H$\alpha$ line measured in the same spectrum. Based on the above, AT~2020wey is a hydrogen and helium (H+He) TDE, but without Bowen lines, a quite rare class as usually H+He TDEs also show Bowen lines. Based on the spectroscopic classification of \citet{Charalampopoulos2022}, it is similar to TDE AT~2017eqx \citep{Nicholl2019}.

We   performed a series of processes before fitting the emission lines of our spectra (see \citealt{Charalampopoulos2022} for a detailed description). The spectra (including the host) were first scaled with the ZTF $g$- and $r$-band photometry, and the host galaxy spectrum was subtracted from the transient plus host spectra. Then we corrected the spectra for galactic extinction and finally we removed the continuum by fitting the emission-free regions of each host-subtracted  spectrum with a fourth-order polynomial. The host-subtracted  and continuum-subtracted spectral series can be seen in the right panel of Fig. \ref{fig:atlas}. After we obtained the emission line spectra, we performed our spectroscopic line study using customized Python scripts employing the \texttt{lmfit} package. We   focus here on H$\alpha$, %as it is the strongest line in the spectra 
and we quantify the line luminosity by integrating under the line profile. In order to measure the velocity width, we used a custom script in Python that first smooths the spectrum and  then locates the data points to  the left and right of the maximum that have flux values closest to   half of the maximum; the script  then calculates the distance between them on the x-axis. We use a custom Monte Carlo method (10\,000 iterations of re-sampling the data assuming Gaussian error distribution) in order to calculate uncertainties for the flux (luminosity) and line width. In the sub-panels of Fig. \ref{fig:ha_vs_sample}, we present the luminosity (left panel) and line width (right panel) and we compare them with those of the sample of \citet{Charalampopoulos2022}. The line luminosity is below $10^{41}\rm\,erg\,s^{-1}$ and  is identical to that of iPTF16fnl. We find that H$\alpha$ peaks with a lag of $\sim8.2\pm 2.8$ days compared to the light curve peak, and in Fig. \ref{fig:Ha_lag_vs_sample} we plot this lag as a function of the peak of the blackbody radius. \citet{Charalampopoulos2022} found that there is a correlation between the two parameters, %(see their Figure 18) 
and AT~2020wey is indeed within 1$\sigma$ from the best linear fit, and its lag is similar to that of iPTF16fnl. The line width of H$\alpha$ is on the order of 10\,000 $\rm km\, s^{-1}$, and reaches the value of $\sim$ 20\,000 $\rm km\, s^{-1}$ only at 8.8d, where the line luminosity peaks. We note that H$\alpha$ has an asymmetric profile, as seen in many previous TDEs (see, e.g., Fig. 17 of \citealt{Charalampopoulos2022}). In the first three spectra, H$\alpha$ seems to show a bump on the blue side of the main peak. Such broad, multiple peak profiles have been seen in some TDEs and, after careful modeling, have been attributed to the formation of highly inclined, highly elliptical, and relatively compact accretion disks (e.g., PTF09-djl; \citealt{Arcavi2014,Liu2017}, AT~2020zso; \citealt{Wevers2022}), consistent with theoretical works predicting that the returning debris should form an extended eccentric accretion flow \citep{Shiokawa2015}. Such modeling is not attempted in our work since the host-subtracted  optical spectra are of low S/N and the transient has faded very quickly in the optical wavelengths (within $\sim$ 20 days after peak).

\begin{figure*}
        \centering
        \begin{subfigure}[b]{1\textwidth}
            \centering
        \includegraphics[trim= 0 -7.5cm 0 0,width=0.492 \textwidth]{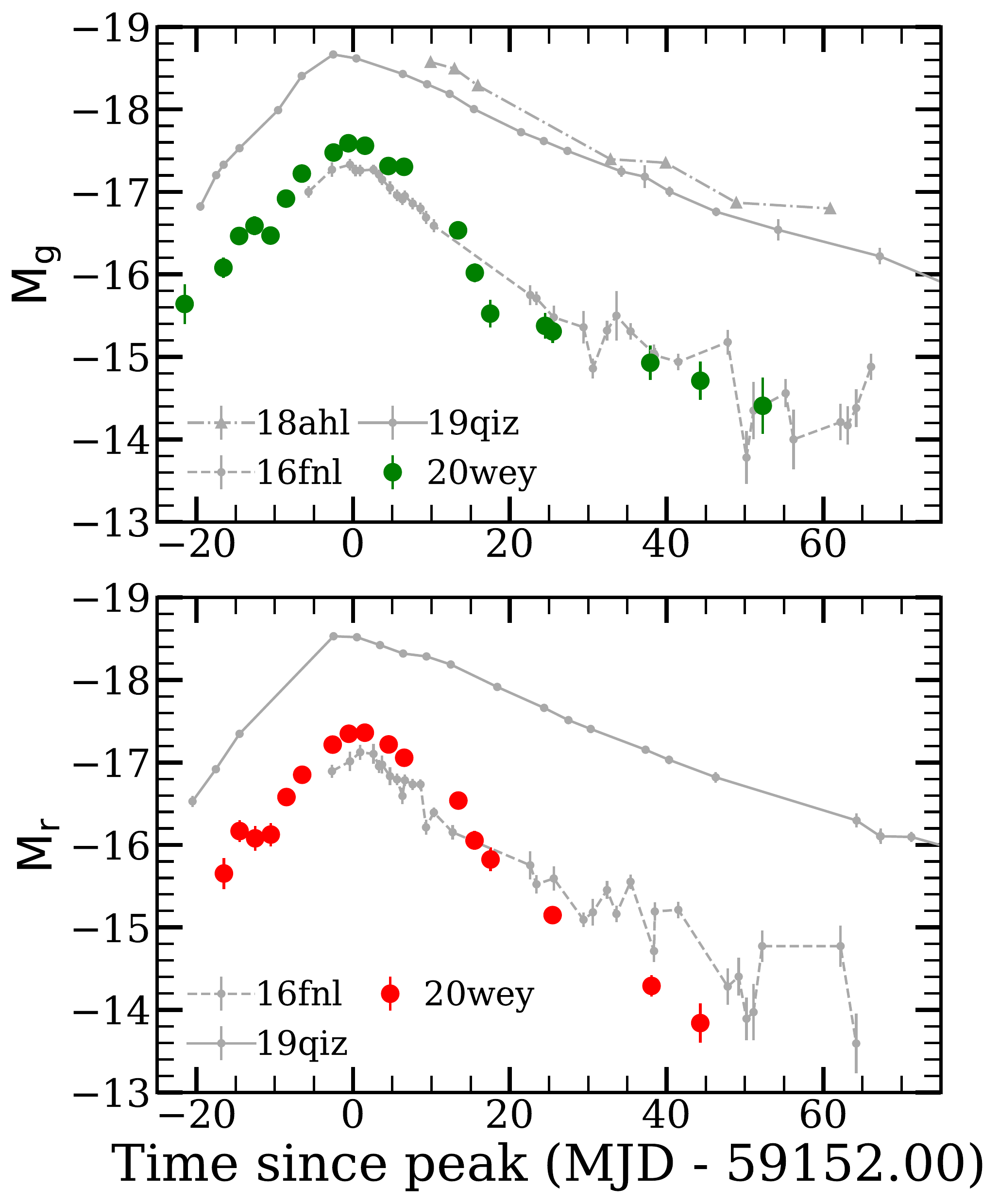}
        \includegraphics[width=0.492 \textwidth]{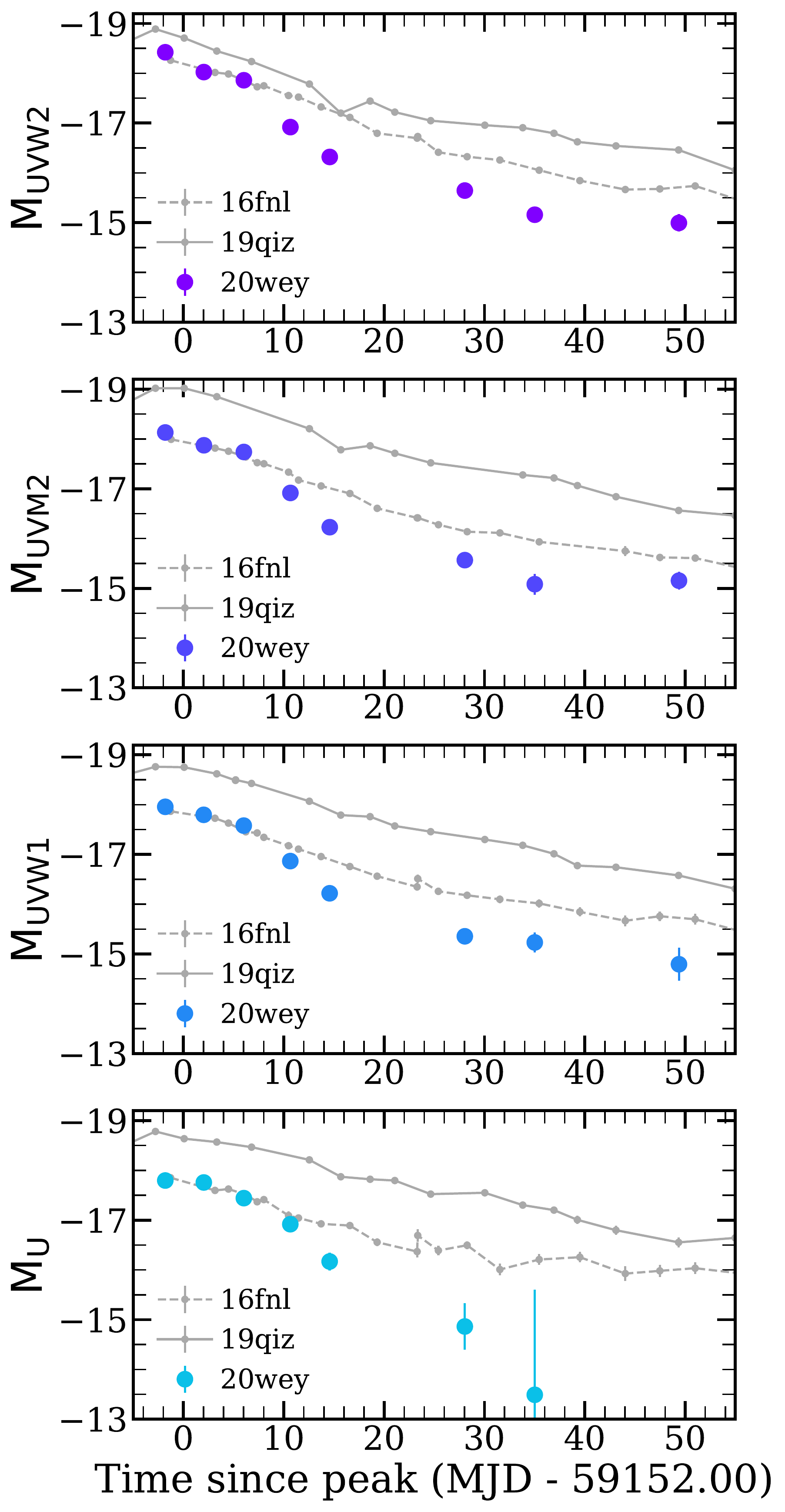}
            % \caption[Network2]%
            % {{\small Network 1}}    
            % \label{fig:off_sub}
        \end{subfigure}
        % \vskip\baselineskip
        \caption{Light curves of AT~2020wey and other faint and fast TDEs. Left panels: Comparison of the ZTF and LCO host-subtracted  and de-reddened magnitudes of AT~2020wey with three faint and fast TDEs (iPTF16fnl, AT~2019qiz, and AT~2018ahl). Right panels: Comparison of the UVOT host-subtracted  and de-reddened magnitudes of AT~2020wey with two faint and fast TDEs (iPTF16fnl and AT~2019qiz). At $+8$ days post-peak, AT~2020wey starts declining very quickly in the UV; at 20 days after peak it has dropped an average of 0.69~mag more in the three UV bands (i.e., UVW1, UVM2, UVW2) compared to iPTF16fnl.}
        \label{fig:comp_ztf_and_uvot}
    \end{figure*}
    
\begin{figure}
  \centering
  \includegraphics[width=0.5 \textwidth]{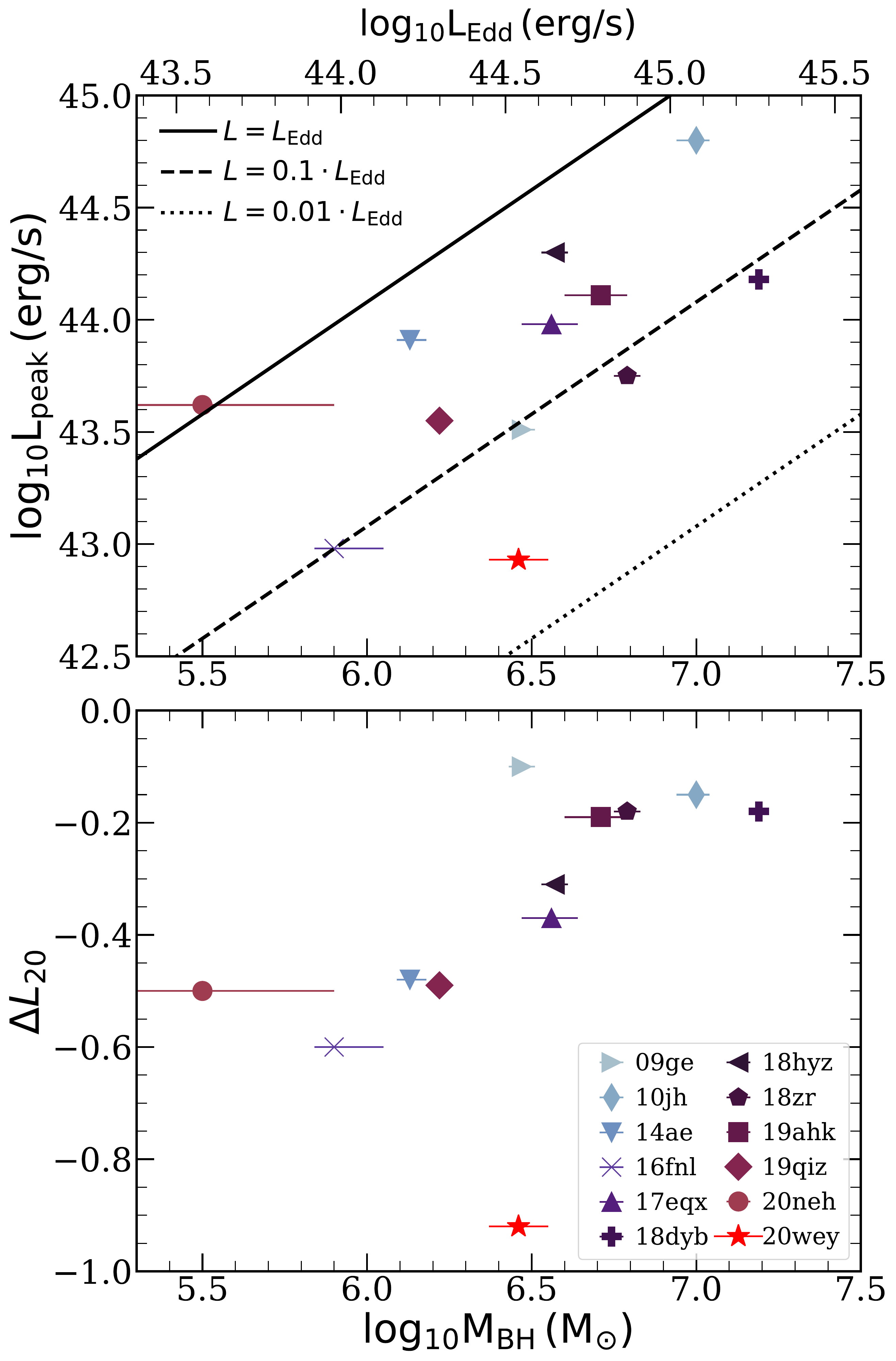}
  \caption{Peak bolometric luminosity (top panel) and its decline rate within the first 20 days ($\Delta L_{20}=\log_{10}(L_{20}/L_{\rm peak}$); bottom panel) as a function of the BH mass, for a sample of optical TDEs. The BH masses are computed with \texttt{MOSFIT} \citep{Mockler2019,Nicholl2022}. There is no correlation between the peak bolometric luminosity and the decline rate. However, there is a strong correlation between the decline rate and the BH mass of TDEs, where AT~2020wey is an outlier. %Using a scaling relation between the BH mass and the total galaxystellar mass, we find a BH mass of ${\log_{10}({\rm M_{BH}/M_{\odot}}) = 6.00 ^{+0.48}_{-0.54}}$. On the other hand, \texttt{TDEmass} \citep{Ryu2020} returns a BH mass of ${\log_{10}({\rm M_{BH}/M_{\odot}}) = 5.61^{+0.07}_{-0.06}}$, almost an order of magnitude lower than \texttt{MOSFIT}.
}
  \label{fig:Mbh_vs_L}
\end{figure}

\section{Discussion} \label{sec:discussion}

\subsection{The nature of faint and fast TDEs} \label{subsec:fandf}

\begin{figure}[h!]
  \centering
  \includegraphics[width=0.5 \textwidth]{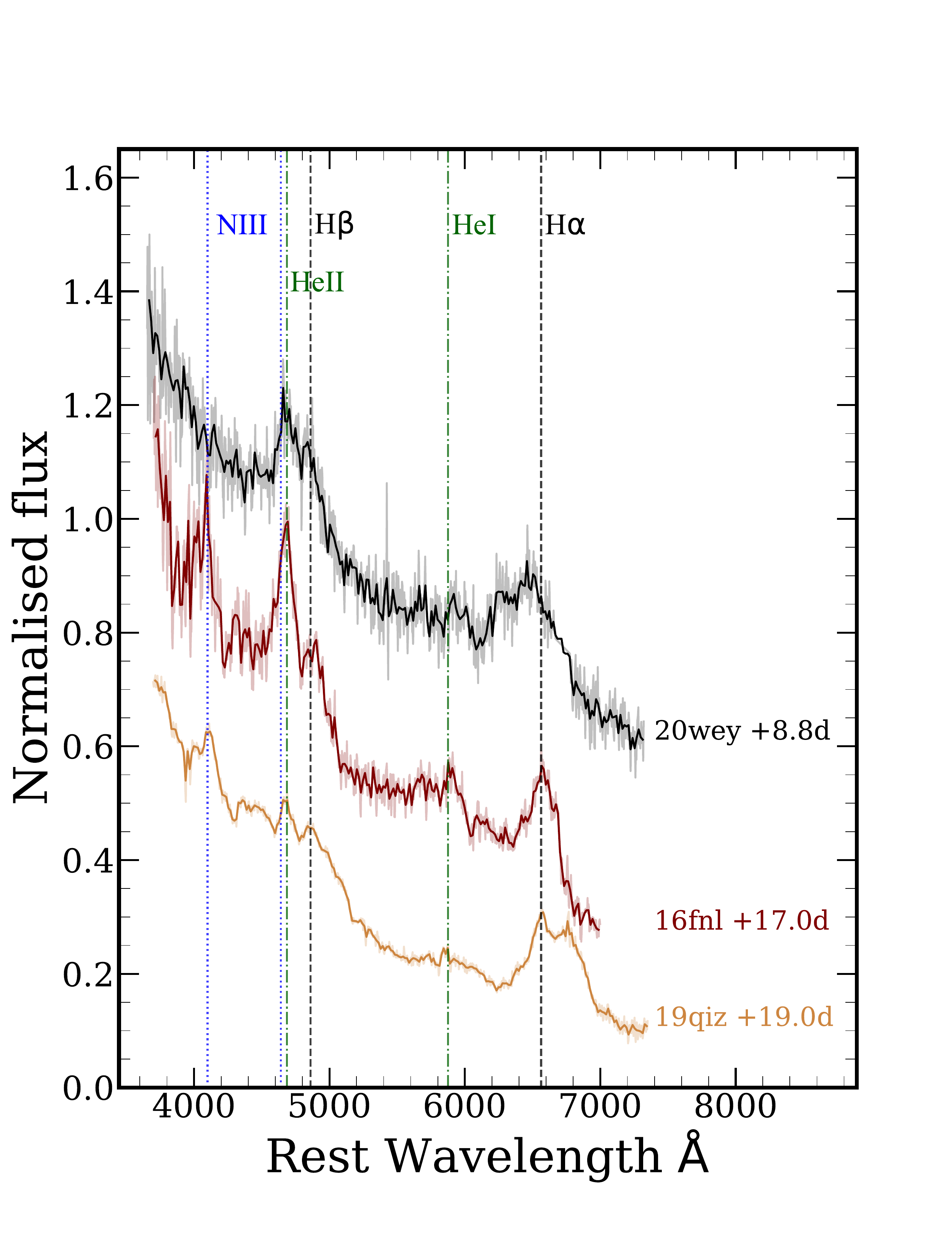}
  \caption{Comparison of the host-subtracted  (and de-reddened) spectra of AT~2020wey with two more faint and fast TDEs (iPTF16fnl and AT~2019qiz).
  The three events are spectroscopically similar, but AT~2020wey lacks the strong nitrogen line at 4100 \AA.} 
  \label{fig:comp_spec}
\end{figure}

The physical origin of what makes a TDE faint and fast remains a mystery to the community. Here, we try to get further insight into this question   by comparing   two well-studied faint and fast TDEs in the literature, iPTF16fnl \citep{Blagorodnova2017,Brown2018,Onori2019} and AT~2019qiz \citep{Nicholl2020,Hung2020a}, as well as the recently studied AT~2018ahl \citep{Hinkle2022} and AT~2020neh \citep{Angus2022}. The following discussion in this section and the comparison of the faint and fast TDEs AT~2020wey, iPTF16fnl, AT~2019qiz, and AT~2020neh is summarized in Table \ref{tab:comp} where we list their properties.

\begin{table}[h]
\renewcommand{\arraystretch}{1.7}
\setlength\tabcolsep{0.01cm}
\fontsize{8.8}{11}\selectfont
\centering
 \caption{Comparison of different properties of AT~2020wey with three well-studied faint and fast TDEs (iPTF16fnl, AT~2019qiz, and AT~2020neh). }
 \label{tab:comp}
 \begin{tabular}{c|c c c c }
  \hline
  & AT~2020wey$^1$ & iPTF16fnl$^2$ & AT~2019qiz$^3$ & AT~2020neh$^4$\\
  \hline
  $^a$$\log({\mh/\msun})$ & - & $5.5^{+0.42}_{-0.42}$ & $^{c}$$(5.75-6.52)^{+0.45}_{-0.45}$ & $4.8^{+0.5}_{-0.9}$ \\
  $^b$$\log({\mh/\msun})$ & $6.46^{+0.09}_{-0.09}$ & $5.9^{+0.15}_{-0.06}$ & $6.22^{+0.04}_{-0.04}$ & $5.5^{+0.40}_{-0.3}$ \\
  $\mstar/\msun$ & $0.11^{+0.05}_{-0.02}$ & $0.98^{+0.03}_{-0.87}$ & $1.01^{+0.03}_{-0.02}$ & $1.3^{+4.90}_{-1.00}$ \\
  b & $1.03^{+0.03}_{-0.08}$ & $0.88^{+0.14}_{-0.03}$ & $0.54^{+0.02}_{-0.02}$ & $1.50^{+0.40}_{-0.90}$ \\
  $L_{\rm X}$ & <6.1 & <0.24 & 5.1 & <45 \\
  ($10^{40}\,\rm erg\, s^{-1}$) & & & & \\
  Spectral type & H+He & H+He+\ion{N}{III} & H+He+\ion{N}{III} & H+He+\ion{N}{III} \\
  Host galaxy & QBS & QBS & Spiral & Dwarf \\
  \hline
  \hline
 \end{tabular}\\
\begin{flushleft}  

% The $\mh$, $\mstar$ and the scaled impact parameter b for each event were retrieved from: $^1$This work (\texttt{MOSFIT}), $^2$\citealt{Nicholl2022} (\texttt{MOSFIT}), $^3$\citealt{Nicholl2022} (\texttt{MOSFIT}), $^4$\citealt{Hinkle2022} ($M-\sigma$).\\
 $^a$ Based on the $M-\sigma$ relation.\\
 $^b$ Based on \texttt{MOSFIT} fit. \\
 $^{c}$ There is a range in the BH masses because three different relations are used \citep{Nicholl2020}.
 \end{flushleft} 
\end{table}

In Fig. \ref{fig:comp_ztf_and_uvot} we compare the broadband light curve evolution, for the ZTF $g$ and $r$ bands in the left panels (optical) and for the UVOT bands W2, M2, W1 and U, in the right panels (UV). We also compare it with AT~2018ahl in the $g$ band alone since there are no observations in $r$ or the UVOT bands. For the comparison in the UVOT bands, we retrieved data from \citet{Hinkle2021a}, who  use the updated calibrations of Swift. The data in this plot are not k-corrected as all those TDEs were found   at comparably low redshifts. In the optical, the decays of the faintest TDEs AT~2020wey and iPTF16fnl look very similar. In the UV, although AT~2020wey and iPTF16fnl have similar magnitudes and evolution until $\sim$ 8 days after the peak, interestingly AT~2020wey starts decaying much faster. At 20 days after peak, AT~2020wey has dropped 0.65--0.73~mag more in the three UV bands (i.e., UVW1, UVM2, UVW2) compared to iPTF16fnl.

\subsubsection{Photometric properties and black hole masses} \label{subsubsec:comp_phot}

In order to make a further comparison in the photometric properties of faint and fast TDEs we retrieved  $\mh$, $\mstar$, and the scaled impact parameter $b$ for iPTF16fnl and AT~2019qiz from \citet{Nicholl2022}, and for AT~2020neh from \citet{Angus2022}, where they fitted the light curves of a sample of TDEs with \texttt{MOSFIT}. We also use the BH mass of AT~2018ahl from \citet{Hinkle2022}, where they estimate it using a stellar mass--black hole mass relation. AT~2020wey and AT~2018ahl have the highest BH masses ($\log({\mh/\msun})\sim6.4$) with iPTF16fnl being the lowest ($\log({\mh/\msun})\sim5.9$) and AT~2019qiz being an intermediate case ($\log({\mh/\msun})\sim6.22$). \citet{Blagorodnova2017}   suggests that the reason for faint and fast TDEs could be the low SMBH mass of the host galaxy.
%, since iPTF16fnl remains the TDE with the lowest estimated BH mass to date. 
However, \citet{Nicholl2020} argues that the estimated mass of AT~2019qiz is not significantly lower compared to the other more slowly evolving TDEs, and that faint and fast events can occur in more massive galaxies. They argue, however, that low BH masses could have an effect on the outflow velocity launched from a TDE since the velocity scales as $v_{\rm esc}\propto M_{BH}^{1/3}$, meaning an outflow can escape more easily at low $M_{BH}$. They also note that the effect could be more prominent in the early phases of the TDE, when the outflow dynamics most likely govern the light curve evolution. \citet{Angus2022} draw the same conclusion for the fastest rising TDE to date, AT~2020neh, for which they measure an intermediate BH mass ($4.8\lesssim\log({\mh/\msun})\lesssim5.9$). If all TDEs have similar Eddington ratios (bolometric-to-Eddington luminosity ratio) as a class, the low BH mass via the early outflow properties could dictate whether a TDE is faint and fast. Here we find that the higher BH mass of AT~2020wey has a slower photospheric radius expansion than AT~2019qiz (see Fig. \ref{fig:randd}) as expected with the above scenario, if the outflow drives the early phase photospheric expansion. However, AT~2020wey then decays more quickly and is also less luminous than AT~2019qiz and AT~2020neh. Here we   note that the BH mass of AT~2020wey was measured with \texttt{MOSFIT} which only provides a model-dependent estimation based on several assumptions. Unfortunately, without a  high-resolution host galaxy spectrum we cannot estimate the BH mass via the  $M-\sigma$ relation. However, we can use the stellar mass of the host galaxy derived from the \texttt{PROSPECTOR} fit (see Sect. \ref{subsec:host_sed}) in order to make an estimation of the BH mass using a scaling relation between the central BH mass and total galaxy stellar mass (similarly to \citealt{Angus2022}) for low-mass galaxies \citep{Reines2015},
\be
\log{\left( \frac{M_{BH}}{M_{\odot}}\right)} = \alpha + \beta \log{\left( \frac{M_{*}}{10^{11}M_{\odot}} \right)}
\label{eq:mbh}
,\ee
where $\alpha = 7.45 \pm 0.08$ and $\beta = 1.05 \pm 0.11$. Using this method we estimate a BH mass of ${\log_{10}({\rm M_{BH}/M_{\odot}}) = 6.00 ^{+0.48}_{-0.54}}$. This value is within 1$\sigma$ of the \texttt{MOSFIT} estimation. Furthermore, in order to make an estimation of the BH mass based on the stream collision scenario \citep{Piran2015,Jiang2016}, we use \texttt{TDEmass} \citep{Ryu2020}, which assumes that the UV--optical emission is powered by shocks occurring due to the debris-stream collision, and extracts the BH and stellar mass based on the observed bolometric luminosity and temperature at the peak. We find a BH mass of ${\log_{10}({\rm M_{BH}/M_{\odot}}) = 5.61^{+0.07}_{-0.06}}$
 and a stellar mass of $ 0.16 ^{+0.04}_{-0.03}$. The  stellar mass is consistent with the \texttt{MOSFIT} estimate, but  the BH mass is not, as \texttt{TDEmass} returns a significantly lower value (almost an order of magnitude lower).

%like it was done for iPTF16fnl \citep{Onori2019} and AT~2019qiz \citep{Nicholl2020} with X-Shooter, however our target was a northern one and we did not have access to a high resolution spectrograph needed for such a task.

In Fig. \ref{fig:Mbh_vs_L} we present the peak bolometric luminosity (top panel) and its decline rate  within the first 20 days ($\Delta L_{20}=\log_{10}(L_{20}/L_{\rm peak}$); bottom panel) as a function of the BH mass, for the sample of optical TDEs presented in Fig. \ref{fig:bol_comp}. The BH masses of the TDEs were computed with \texttt{MOSFIT} and retrieved from \citet{Nicholl2022} (and from \citealt{Angus2022} for AT~2020neh). There  is no correlation between the black hole mass and the peak luminosity (upper panel), which is consistent with previous studies \citep{BLANCHARD2017,Hung2017,Wevers2019a}. In the same panel we show the Eddington ratios (bolometric-to-Eddington luminosity ratio). An interesting feature of AT~2020wey is that it stands out as the one with the lowest Eddington ratio of the sample.

However, there is a strong correlation between the \texttt{MOSFIT}-derived BH masses of TDEs and their decline rate even though AT~2020wey is an outlier in this strong trend. We ran a Pearson's correlation coefficient test which measures the linear correlation between two sets of data. In order to ensure that any measured correlation is not driven by a few extreme data points, we applied a bootstrapping method where, within each iteration (out of $10^5$), we randomly draw N data points with replacement (where N is the number of data points for which we want to measure the correlation for). Without AT~2020wey, the score of Pearson's $r$ test is $0.81\pm^{0.08}_{0.09}$, indicating a very strong correlation ($p$ value of $0.009\pm^{0.002}_{0.007}$). If we include AT~2020wey the score drops to $0.61\pm^{0.15}_{0.24}$, as expected since Pearson's $r$ test is sensitive to outliers ($p$ value of $0.05\pm^{0.04}_{0.11}$).  Spearman's $\rho$ test  for monotonicity is less sensitive to outliers, and  results in a score of $0.75\pm^{0.14}_{0.16}$ ($0.77\pm^{0.22}_{0.16}$  without AT~2020wey). Hence we show that there is a significant ($p\leq 0.05$) linear (and monotonic) correlation between the BH mass and the luminosity decline rate of TDEs. We note that this correlation might be expected as \texttt{MOSFIT} estimates the BH mass exactly through fitting the light curves and assuming fallback, whose rate is dependant on the BH mass.
%hence it provides estimates that could be sensitive to the light curve slope and its evolution. 
What is important, however, is that AT~2020wey appears to be an outlier in this strong correlation. 
This could indicate that its fast decline from the peak, may be driven by different physical processes than in other TDEs or than those assumed in the \texttt{MOSFIT} model. Even if this is  so, the posteriors we derived for the BH and the stellar mass should still stand since the uncertainty on the driving mechanism is encapsulated in the efficiency parameter $\epsilon$, as it is possible that quite different mechanisms (shock heating or accretion) could  track the mass return rate, just with different normalizations (i.e., efficiencies).  
In the Appendix we provide a plot similar to Fig. \ref{fig:Mbh_vs_L}, with the difference that the BH masses have been estimated through the M--$\sigma$ relation
\citep{Wevers2019,Leloudas2019,Short2020a,Nicholl2020,Angus2022}. 
Unfortunately, this   estimate does not exist for AT~2020wey, and it is therefore not included in the plot. 
%For this reason, in order to further confirm the aforementioned correlation, we also compare with BH mass estimates of the above sample using two other methods, presented in the left and right panels of Fig. \ref{fig:Mbh_corr} in the Appendix. The first is the M-$\sigma$ relation where we retrieve the (available) data from \citet{Wevers2019a} or from the individual study paper of each event (\citealt{Leloudas2019,Short2020a,Nicholl2020,Angus2022}).
 Although weaker (Pearson's $r$ test result of $0.69\pm^{0.11}_{0.16}$ with a $p$ value of $0.05\pm^{0.03}_{0.09}$), the correlation between BH mass and light curve decline rate still exists, showing that it is not uniquely driven by the \texttt{MOSFIT} systematics.

\citet{Hinkle2021a} reports a correlation between the peak value of the bolometric luminosity and the the decline rate within the first 40 days ($\Delta L_{40}=\log_{10}(L_{40}/L_{\rm peak}$)) in optical TDEs and derives an empirical relation between the two quantities (see their Equation 1). We measure the $\Delta L_{40}$ of AT~2020wey and find it to be $-1.28$, within $1\sigma$ of the expected value derived from their empirical relation (i.e., ${-1.21^{+0.17}_{-0.18}}$). Therefore, AT~2020wey is consistent with the correlation proposed by \citet{Hinkle2021a} at the low-luminosity end. %confirms the correlation of $L_{\rm pk}$ and $\Delta L_{40}$.

\subsubsection{Other observables and spectroscopic properties} \label{subsubsec:comp_rest}

 In this section we investigate whether there is any single observable property that is common among faint TDEs. \texttt{MOSFIT} suggests that iPTF16fnl and AT~2019qiz were caused by the partial disruption of solar mass stars, while AT~2020wey  by the complete disruption of a sub-solar mass star ($\mstar/\msun=0.11^{+0.05}_{-0.02}$). We also investigated if the faint and fast TDEs were similar regarding their X-ray properties, but we could not identify a clear tendency; AT~2019qiz was detected in X-rays, while the others were not detected (or were marginally detected). AT~2019qiz had some time-resolved detections in X-rays and was measured in a single deep stack image to have a luminosity of $\sim 5.1\times10^{40}\,\rm erg\, s^{-1}$ in the 0.3–10~keV band. A single deep stack image in the same energy band for iPTF16fnl provided an extremely weak marginal detection of $\sim 2.4\times10^{39}\,\rm erg\, s^{-1}$. We also consider that the host galaxies of the events and AT~2019qiz occurred in a spiral galaxy that probably harbored a weak AGN; however, iPTF16fnl and AT~2018ahl occurred in quiescent Balmer-strong galaxies, while AT~2020neh occurred in a dwarf galaxy. 
%However, even if all of them occurred in such galaxies, we would not be able to robustly associate it with faint and fast TDEs, since 32\% of (optical) TDEs have been discovered in such galaxies \citep{French2020}.

We finally compare the faint and fast TDEs based on their spectroscopic properties. Fig. \ref{fig:comp_spec} shows the host-subtracted  and de-reddened spectra of AT~2020wey ($+8.8\rm d$) (our best S/N spectrum) with iPTF16fnl ($+17.0\rm d$) and AT~2019qiz ($+19.0\rm d$). The spectra of the three events are very similar. We see broad and asymmetric H$\alpha$; we see that the peak of H$\beta$ is nicely resolved in the $\sim$4600\AA\, blend; and there is  evidence for weak \ion{He}{I} in all of them. All three events have a strong peak at \ion{He}{II} 4686 \AA,\, which may be blended with \ion{N}{III} 4640 \AA. The reason why iPTF16fnl and AT~2019qiz are classified as \ion{N}{III} rich TDEs (Bowen) is that they show a clear strong peak of \ion{N}{III} $\sim$ 4100\AA,\, which is absent for AT~2020wey in all of our spectra. Hence, we can robustly classify AT~2020wey as a H+He TDE with no Bowen features. This means that there does not seem to be a connection between faint and fast TDEs and the physical origin of Bowen fluorescence lines. 

The faint and fast TDEs appear to be different  in many observables, and we conclude that none of these properties shows a tendency for the four faint and fast TDEs.

\begin{figure*}
        \centering
        \begin{subfigure}[b]{1\textwidth}
            \centering
        \includegraphics[trim={0 0cm 0 0cm},clip,width=0.497 \textwidth]{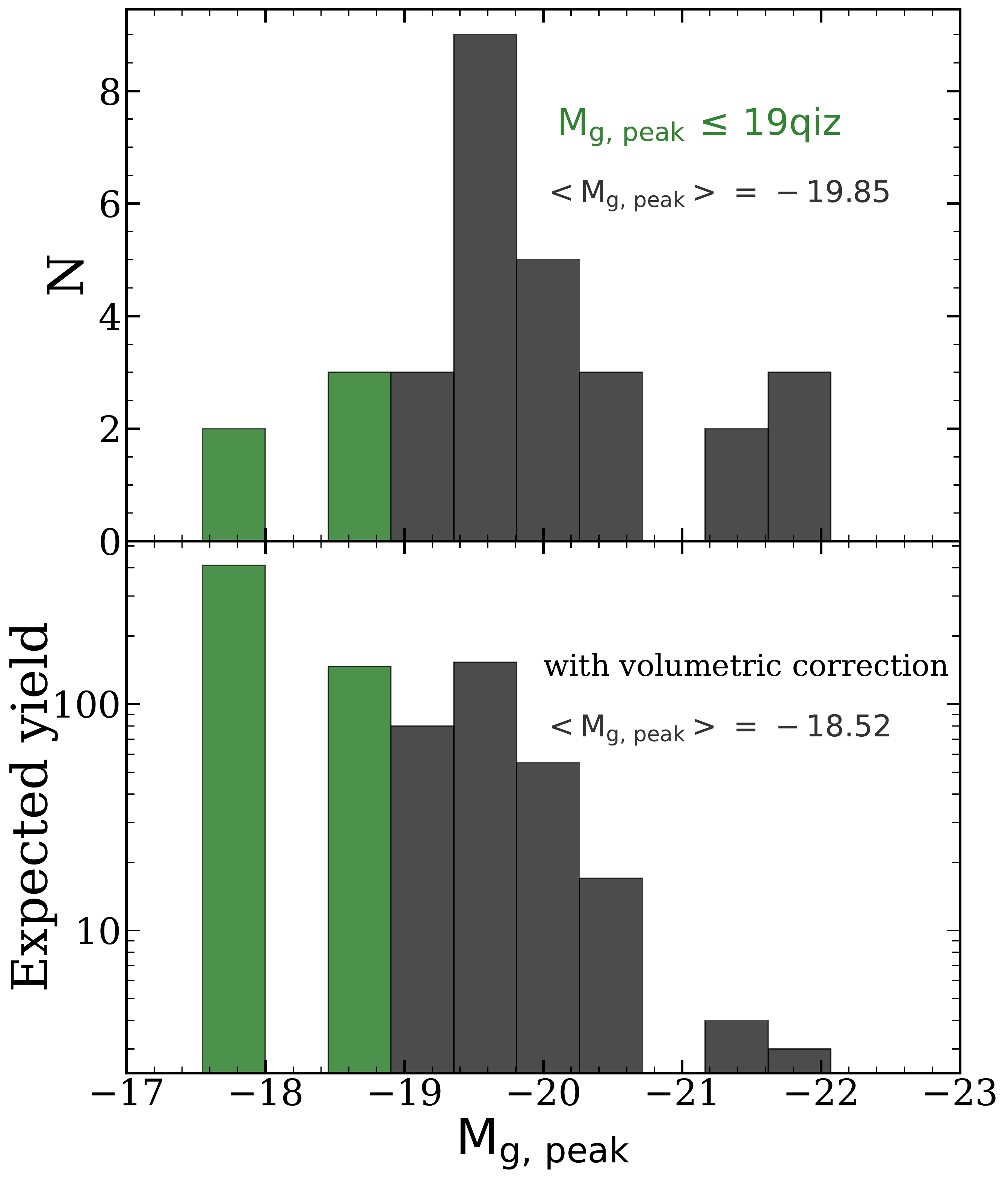}
        \includegraphics[trim={0 0cm 0cm 0},clip,width=0.497 \textwidth]{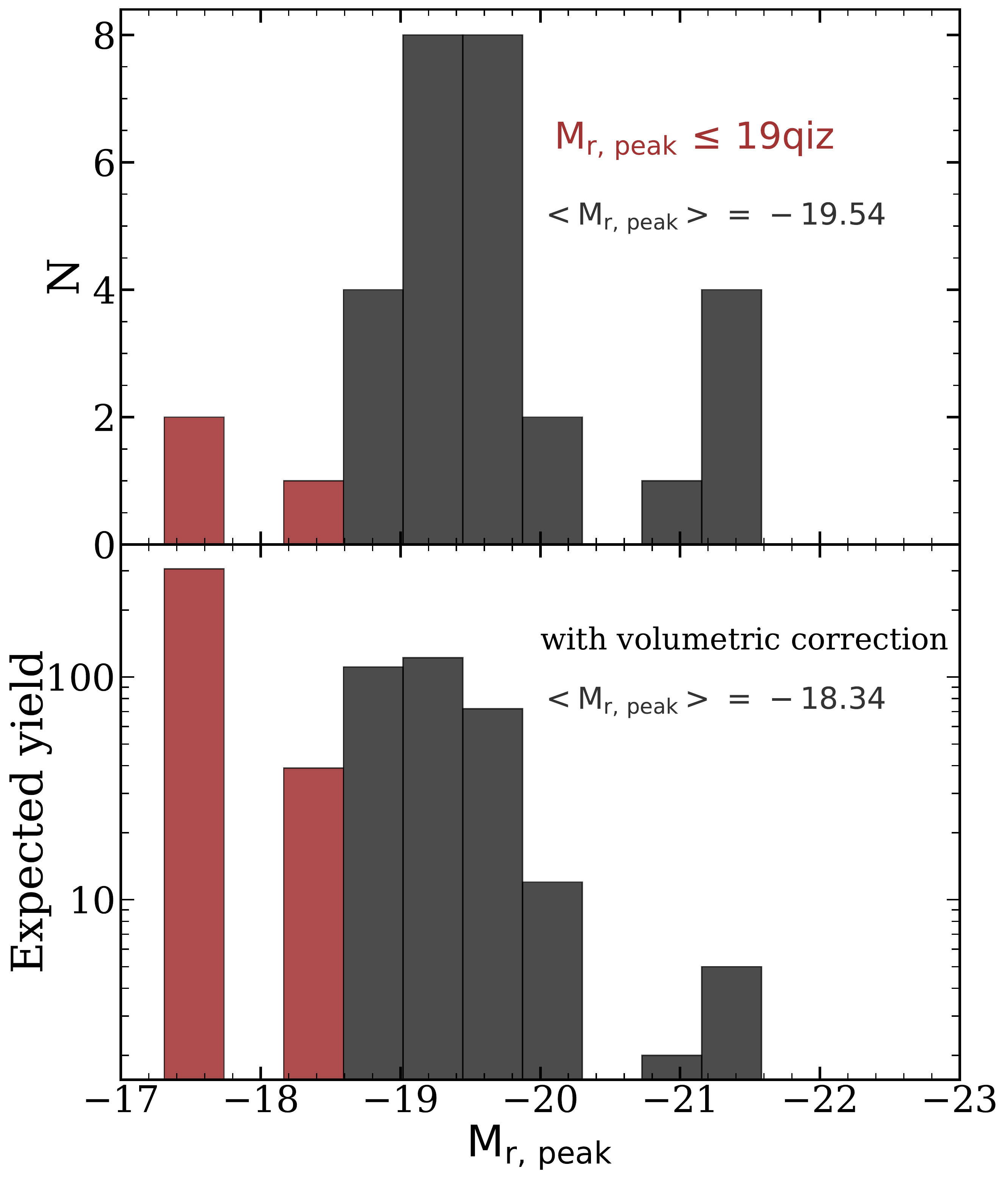}
            % \caption[Network2]%
            % {{\small Network 1}}    
            % \label{fig:off_sub}
        \end{subfigure}
        % \vskip\baselineskip
        \caption{Absolute peak magnitude distributions (k-corrected) of the \citet{Hammerstein2022} TDE sample, with $g$ band on the left and $r$ band on the right. Shown in  the top panel are the histograms of the original data (observed distributions) and in the bottom panel the data after volumetric correction (i.e., expected yields, number of TDEs). The colored bins   contain TDEs that are fainter than or equally faint to AT~2019qiz.}
        \label{fig:vol_cor_hists}
    \end{figure*}

\subsection{Luminosity function and TDE rates} \label{subsec:rates}

In order to further probe the nature of faint TDEs, we investigate whether they are indeed rare by nature or if we have been biased toward bright nuclear flares (peak absolutes magnitudes of $\sim$ -20 mag) and and have caused fainter TDEs to go unnoticed in searches within the new detections of wide-field surveys. That is, we want to account for observational biases, for example  the Malmquist bias, which describes the preferential detection of intrinsically brighter objects in brightness-limited surveys (\citealt{Malmquist1922}) and quantify how many  faint TDEs are missed.

\begin{figure*}
        \centering
        \begin{subfigure}[b]{1\textwidth}
            \centering
        \includegraphics[width=0.480 \textwidth]{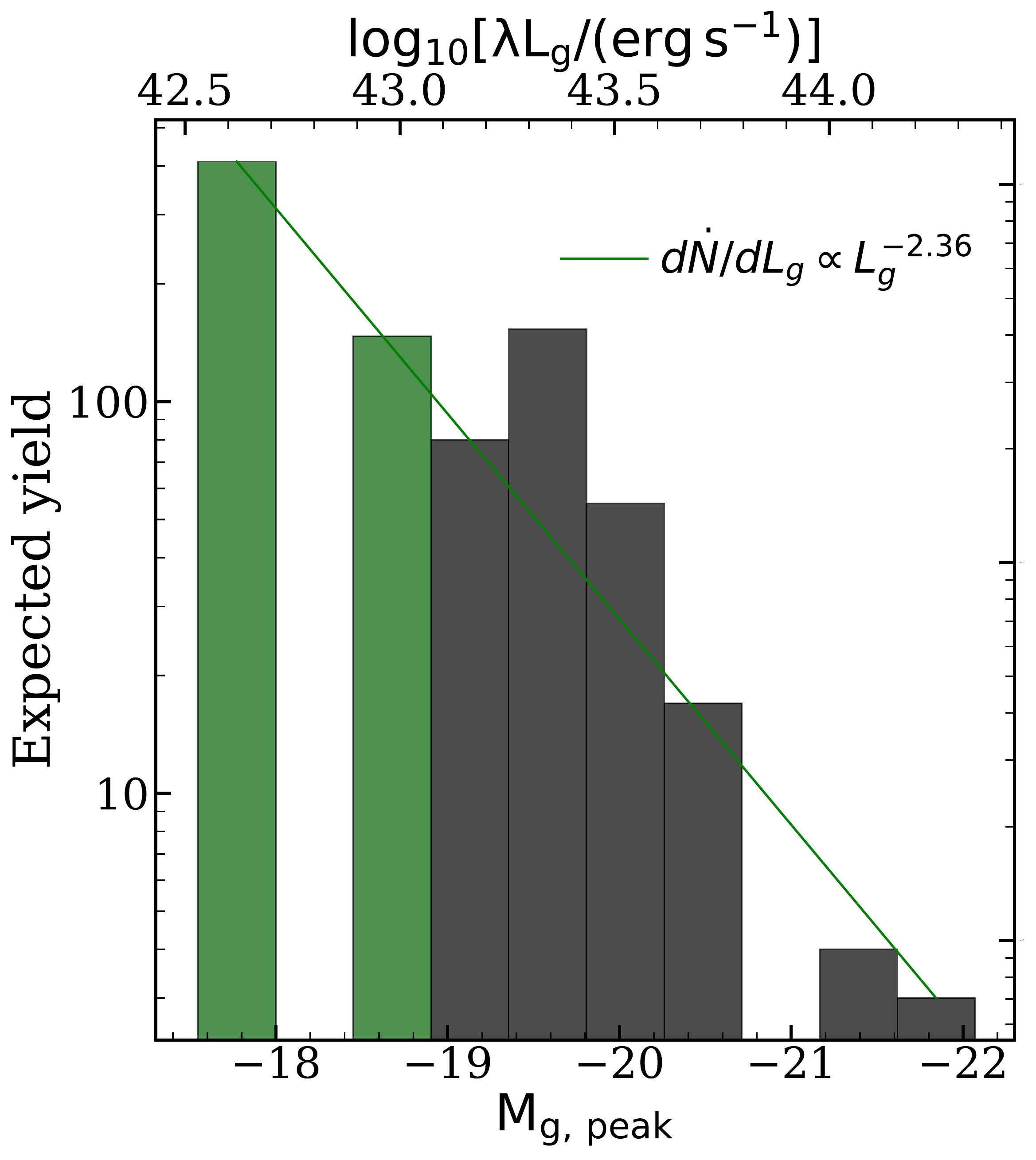}
        \includegraphics[width=0.504 \textwidth]{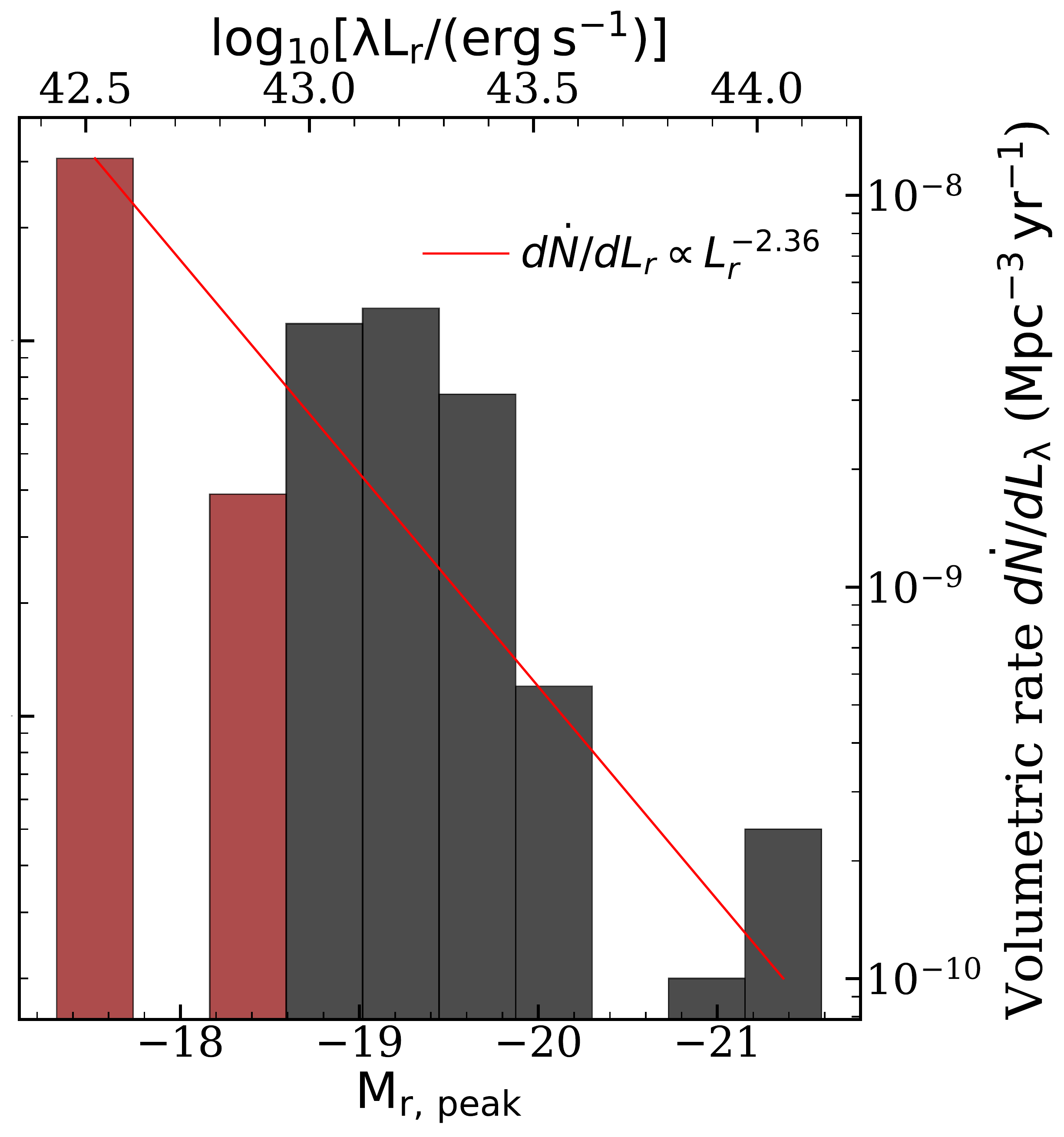}
            % \caption[Network2]%
            % {{\small Network 1}}    
            % \label{fig:off_sub}
        \end{subfigure}
        % \vskip\baselineskip
        \caption{Fits to the volumetric rates based on the \citet{Hammerstein2022} sample, for the ZTF $g$ band (left) and ZTF $r$ band (right). The histograms denote the expected yields (number of TDEs) after the volumetric correction, and they are the same as the bottom panel of Fig. \ref{fig:vol_cor_hists}. The colored bins include TDEs that are fainter than or equally faint to AT~2019qiz.}
        \label{fig:lf_fits}
    \end{figure*}

For this reason, we use the same sample presented by \citet{Hammerstein2022}, which consists of 30 spectroscopically classified TDEs observed by ZTF Phase-I survey (see further details and their selection criteria therein). This sample includes two well-studied faint and fast TDEs: AT~2020wey and AT~2019qiz.
% This sample has the advantage that it includes both 2020wey and 2019qiz. 
We first took the public ZTF difference photometry of those TDEs at peak, and isolated the peak of each event in the  $g$ and $r$ bands. Then we computed the absolute magnitudes after correcting for galactic extinction  (host galaxy extinction is not considered) and performing a k-correction. A k-correction is an important correction for such samples because of the vast redshift interval between the different events. We computed a 0.98 magnitude correction in the $r$ band for the highest redshift event in the sample (AT~2020riz; z=0.435). We performed the k-correction using the peak blackbody temperatures of each TDE as estimated by \citet{Hammerstein2022}. 
% To perform a K-correction, we assumed that the SED follows a Planck function and we used the blackbody temeperatures from H22
In the top panel of Fig. \ref{fig:vol_cor_hists} we show the peak luminosity distribution of the 30 TDEs in \citet{Hammerstein2022}, in the ZTF $g$ (left) and $r$ (right) bands. In order to visualize faint TDEs we set an arbitrary cut for events being fainter than or equally faint to AT~2019qiz and we plot those histogram bins in green for the $g$ band and in red for the $r$ band. We find an average peak of $<M_{g,\,\rm peak}>=-19.85$~mag for the $g$ band and $<M_{r,\,\rm peak}>=-19.54$~mag for the $r$ band. 

However, in order to make a fair comparison we need to properly address observational biases. The Malmquist bias is taken account of by normalizing the numbers of each bin to the same co-moving volume. We calculate the volumetric correction $V_c$ for each TDE as the ratio of the volume $V_{\rm max}$ probed by the most luminous TDE in the \citet{Hammerstein2022} sample (AT~2018jbv with $M_{g,\,\rm max}=-22.07$~mag and $M_{r,\,\rm max}=-21.58$~mag) to the volume $V_{\rm max, i}$ probed by each individual TDE, if they were observed at a brightness equal to the faintest apparent magnitude of the sample (AT~2020ocn with $m_{g,\,\rm min}=19.65$~mag and $m_{r,\,\rm min}=19.79$~mag). The redshift at which each TDE would have been observed with this apparent magnitude is defined as $z_{\rm i}$, with $z_{\rm max}$ being the highest of them all (attributed to AT~2018jbv). The volumetric correction factor $V_{\rm c,i}$ for each TDE (see, e.g., \citealt{DeCia2018}) can be written as 
\begin{equation} \label{eq:vc}
V_{\rm c,i} = V_{\rm max} / V_{\rm max, i}  =   \left( \frac{D_{L, \rm max}}{1+z_{\rm max}}\right) ^3 /  \left( \frac{D_{L, \rm max, i}}{1+z_i}\right) ^3 \mbox{,}
\end{equation}
where the luminosity distance of the brightest TDE in the sample (in each respective band $\lambda$) is $D_{L, \rm max} = 10^{(((m_{\lambda,\,\rm min} - M_{\lambda, \rm max}) +5)/5)}$, and the luminosity distance at which each individual TDE would have been observed with the faintest apparent magnitude is $D_{L, \rm max, i} =10^{(((m_{\lambda,\,\rm min} - M_{\lambda, i}) +5)/5)}$. The bottom panel of Fig. \ref{fig:vol_cor_hists} shows the peak magnitude distribution after applying the volumetric correction. We find that %after applying the volumetric correction (hence correcting for observational biases), 
the mean peak absolute magnitudes of the sample reduces to $<M_{g,\,\rm peak}>=-18.52$~mag and $<M_{r,\,\rm peak}>=-18.34$~mag,  which implies a magnitude difference from the observed sample of 1.33~mag and 1.2~mag in $g$ and $r$ band, respectively. These large differences suggest that we are indeed biased toward finding bright nuclear flares, and that faint events like AT~2020wey are not as rare as they are considered to be. The events that are fainter than or  as faint as AT~2019qiz in the observed sample constitute barely 10\% in the $g$ band (16\% in $r$ band) but they are estimated to be 52\% (64\%) after the volumetric correction.
The fact that faint TDEs should constitute up to (50-60) \% of the whole population could alleviate some of the tension between the observed and theoretical TDE rate estimates (about  an order of magnitude discrepancy; see, e.g., review of \citealt{Stone2020}).

% Following \citet{Velzen2018} and \citet{Hung2018}, we also calculate the volumetric rate of TDEs. 

Based on our volumetric correction of the \citet{Hammerstein2022} sample, we calculate the volumetric rate of TDEs. The number of TDEs detected by the survey (see, e.g., \citealt{Hung2018}) can be expressed as

\begin{equation}
\label{eq:rate}
N_{TDE} = \int \frac{d\dot{N}}{dL_{\lambda}} \times \frac{4\pi}{3}D_{L,max}^3 \times A_{\rm survey} \times \tau_{\rm survey}~dL_{\lambda} \equiv \sum_{i=1}^{N} V_{\rm c,i}
,\end{equation}
where $d\dot{N}/{dL_{\lambda}}$ is the luminosity function of TDEs (the volumetric TDE rate with respect to peak $g$-band luminosity), $D_{L, \rm max}$ is the maximum distance (redshift) within which a TDE can be detected by the flux-limited survey (and is a function of the peak magnitude of the flare), and $A_{\rm survey}\times \tau_{\rm survey}$ is used to label the product of the effective survey area (ZTF areal coverage of $23\,675\,deg^{2}$ over the total area of the celestial sphere) and duration. For the duration $\tau_{\rm survey}$, we set a starting date at 1 October 2018 and an end date at 1 December 2020. The \citet{Hammerstein2022} sample consists of 30 spectroscopically classified TDEs observed by ZTF during the ZTF Phase-I survey. However, we see that while AT~2018zr was observed in March 2018 and AT~2018bsi in April 2018, the next TDE is observed on 4 October 2018 (i.e., $\sim$ six months later) and there is a regular cadence in TDE observations from then on. Hence, we exclude the first two events from our calculations and we consider the start of the survey that observed 28 TDEs to be   1 October 2018. All these assumptions, combined with the selection criteria  based on which a discovered transient was spectroscopically classified (i.e., some TDEs during ZTF Phase-I might never got classified, and \citealt{Hammerstein2022} do not claim that their sample is spectroscopically complete), make our estimate of the TDE volumetric rate a rough lower limit.

On the primary y-axis (left) of Fig. \ref{fig:lf_fits}, we show the expected yield,   the expected number of TDEs observed by the survey within each magnitude bin range (same ranges as in Fig. \ref{fig:vol_cor_hists}). On the secondary y-axis (right) we present the volumetric rate $d\dot{N}/{dL_{\lambda}}$ binned by the peak absolute magnitude, for the $g$ band in the left panel and for the $r$ band in the right panel. We fit the volumetric rate in each band with a power law and we get the same power-law index for both bands: ${b=-2.36\pm0.16}$ for $g$ band and ${b=-2.36\pm0.31}$ for $r$ band. Hence, we find that the luminosity function of TDEs can be described by ${d\dot{N}/{dL_{\lambda}}\propto  L^{-2.36}}$ and we conclude, similarly to \citet{Velzen2018}, that a steep power law is required to explain the observed rest-frame luminosities of TDEs (they find ${d\dot{N}/{dL_g}\propto L^{-2.5}}$). When our work was at a late stage, a thorough study on the luminosity function of TDEs from the ZTF-I survey was published by \citet{Lin2022}, who studied almost the same sample as ours. In their work, they find a luminosity function of $dN/dL_{g} \propto {L_{g}}^{-2.3\pm0.2}$, practically the same as we do.

Finally, if we sum the contributions of each volumetric rate bin (see Fig. \ref{fig:lf_fits}) we can derive a lower limit  for the TDE rate, and we calculate it to be $\dot{N}\sim2\times 10^{-8}$ Mpc$^{-3}$ yr$^{-1}$. \citet{Hung2018}   calculates a lower limit for TDEs in red galaxies of $\dot{N}\sim1.1^{+1.8}_{-0.8}\times 10^{-7}$ Mpc$^{-3}$ yr$^{-1}$; this value is consistent with ours since their study focuses on TDEs in red galaxies (easier to detect since TDEs have intrinsically blue colors). The two values would most likely be within one sigma of each other if we could calculate error bars on our measurement, but we cannot since the largest source of uncertainty is systematic (whether the \citealt{Hammerstein2022} sample is spectroscopically complete or not) and was not controlled by us. \citet{Lin2022} calculated a TDE rate of $\dot{N}\sim6.3\times 10^{-8}$ Mpc$^{-3}$ yr$^{-1}$, that is $\sim$ three times larger than ours. A possible explanation for the small discrepancy is that they used a limiting magnitude of $m_{r}=19.5$~mag, while ours was set by the faintest TDE in the sample ( $m_{r,\,\rm min}=19.79$~mag). Another possible reason is that they included the two early TDEs of the ZTF-I survey that we did not, and also   included three more than those in the \citealt{Hammerstein2022} sample (AT2020neh, AT2020nov, AT2020vwl), setting the effective survey duration to 2.7 years (compared to our 2.16 years). We   made a calculation including the two early TDEs (raising our sample to 30), and thus extending the effective survey duration, as well as setting our limiting magnitude to 19.5. In this way we find a TDE rate of $\dot{N}\sim3.5\times 10^{-8}$ Mpc$^{-3}$ yr$^{-1}$, which is 1.8 times smaller than   that  found by \citet{Lin2022}. Hence, we conclude that adding those  three extra TDEs is most likely the reason for this small discrepancy. \citet{Velzen2018} found a rate of $(8\pm 4) \times 10^{-7}\, {\rm Mpc}^{-3}{\rm yr}^{-1}$; here the discrepancy is very large and more extensive. A thorough analysis on why this discrepancy exists can be found in the recent work of \citet{Lin2022}. However the authors cannot conclude on the reason for their lower volumetric rate compared to the one of \citet{Velzen2018}. 

\begin{figure}[h!]
  \centering
  \includegraphics[width=0.5 \textwidth]{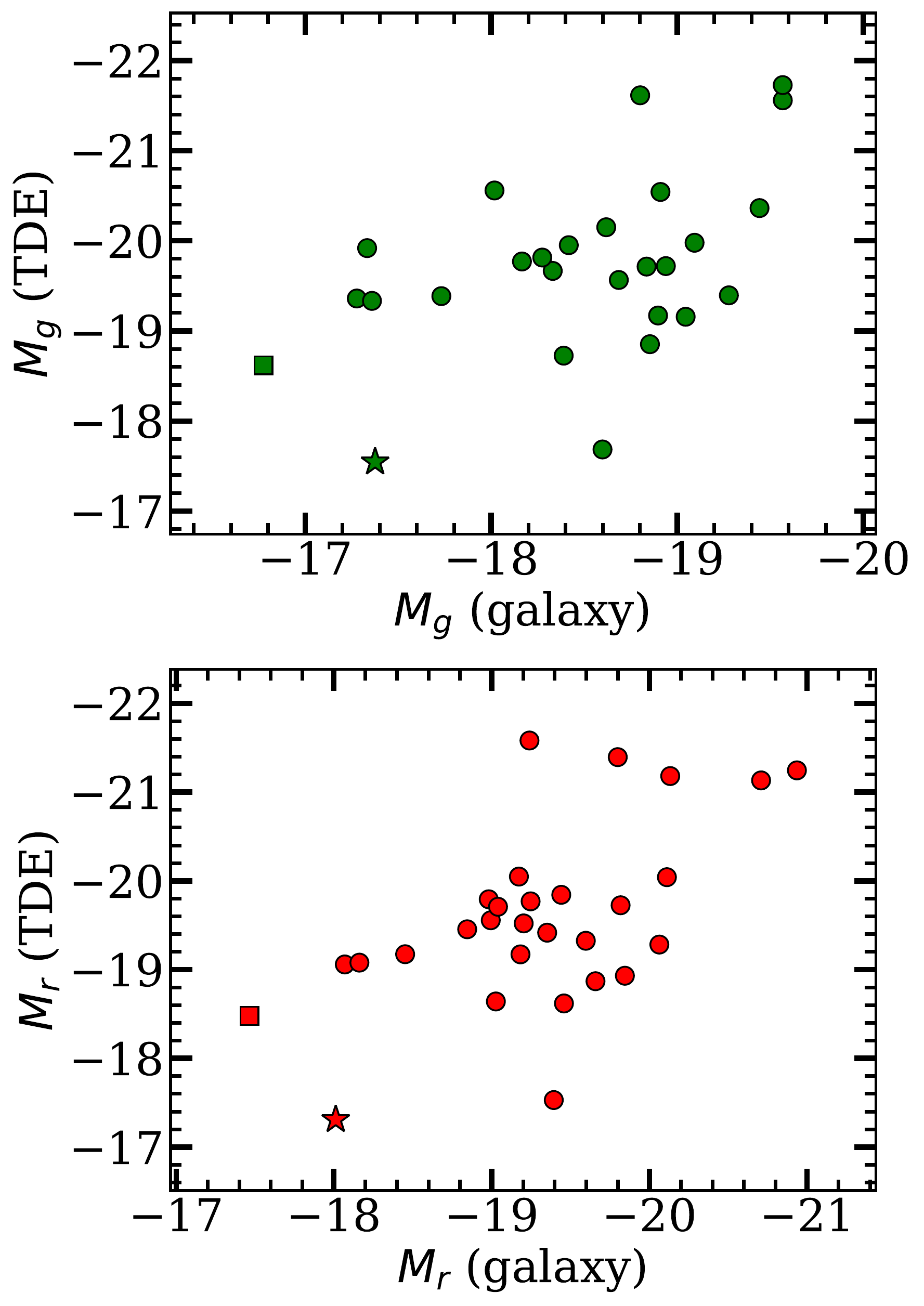}
  \caption{Absolute magnitudes of the TDEs in the \citet{Hammerstein2022} sample against the absolute magnitude of the nucleus of their respective hosts, in the ZTF $g$ (top panel) and $r$ (bottom panel) bands. AT~2020wey is plotted as a star and AT~2019qiz as a square. Even though the two TDEs   have a faint host nucleus compared to the rest of the sample, a bias was not found to be present    in discovering faint TDEs because of the faint nuclei of the  respective hosts. } 
  \label{fig:gal_comp}
\end{figure}

Finally, we calculate the rate of faint TDEs (defined as being as faint as or fainter than AT~2019qiz) to be $\dot{N}\sim 10^{-8}$ Mpc$^{-3}$ yr$^{-1}$ making faint TDEs half of the entire TDE population. In order to investigate whether we might be biased in finding faint TDEs in faint-nucleus galaxies, in Fig. \ref{fig:gal_comp} we plot the absolute magnitudes of the TDEs in the \citet{Hammerstein2022} sample against the absolute magnitude of the nucleus of their respective hosts (retrieved from the \texttt{Lasair} alert broker), in the ZTF $g$ and $r$ bands. We do not find a strong correlation between the two quantities, meaning that we are not actually biased in discovering faint TDEs because of the faint nuclei of their respective hosts. However,   AT~2020wey and AT~2019qiz both have a faint host nucleus compared to the rest of the sample, making them easier targets to initiate a follow-up campaign for.

\section{Conclusions} \label{sec:conclusion}

We   presented the follow-up and detailed study of AT~2020wey, an optical--UV TDE candidate in a quiescent, Balmer-strong galaxy at 124.3 Mpc, along with a comprehensive analysis of its photometric and spectroscopic properties and evolution. We summarize our key findings here:

\begin{enumerate}[label={\arabic*.}]

\item AT~2020wey is a very faint TDE with a peak absolute magnitude of $M_{g} = -17.45$ mag and a peak bolometric luminosity of $L_{\rm pk}=(8.74\pm0.69)\times10^{42} \rm\,erg\,s^{-1}$, making it equally faint to iPTF16fnl, the faintest TDE to date.

\item AT~2020wey is a fast-evolving TDE. The time from the last non-detection to the $g$-band peak is (23 $\pm$ 2) days, and the decline rate within the first 20 days after peak ($\Delta L_{20}=\log_{10}(L_{20}/L_{\rm peak}$) is the fastest measured in any TDE. The broadband luminosity decline during the first 20 days is linear with time.

\item The bolometric light curve analysis shows a blackbody temperature of $\tbb\sim20\,000\rm\,K$ around peak followed by a slow rise, and a blackbody photosphere that expands with a constant velocity ($\sim 1\,300\rm\,km\, s^{-1}$) to a peak value of $\rbb\sim3.5\times10^{14}\rm\,cm$ before it starts shrinking.

\item The early optical emission shows a luminosity evolution $L\propto t^{1.8}$ (consistent with other TDE pre-peak evolution timescales) followed by a sharp fading (best described as an exponential decay) that is steeper than the canonical $t^{-5/3}$ as well as other fast TDEs. This makes AT~2020wey the fastest decaying TDE to date.

\item Our spectroscopic dataset reveals broad ($\sim10^{4}\,\rm km\,s^{-1}$) Balmer and \ion{He}{II} 4686 \AA\, lines and that H$\alpha$ peaks with a lag of $\sim8.2\pm2.8\rm\,d$ compared to the light curve maximum light. The spectrum of AT~2020wey is very similar to those of iPTF16fnl and AT~2019qiz (around $+10\rm d$ post-peak) with the exception that AT~2020wey does not show any \ion{N}{III} Bowen lines. The lack of obvious Bowen lines suggests that there is no connection between the physical origin of Bowen fluorescence lines and faint and fast TDEs.

\item{After making a comparison between the properties of a sample of faint and fast TDEs,  we conclude that there is not a single observable that is common between all of them.}

\item There is a strong correlation between the \texttt{MOSFIT}-derived BH masses of TDEs and their decline rate within the first 20 days ($\Delta L_{20}=\log_{10}(L_{20}/L_{\rm peak}$). However, AT~2020wey is an outlier in this correlation, possibly indicating that its fast early decline could be dictated by a different physical mechanism than fallback.

\item After performing a volumetric correction to a sample of 30 TDEs observed between 2018 and 2020 \citep{Hammerstein2022}, we conclude that faint TDEs are not rare by nature and that they should constitute up to $\sim$(50-60)\% of the entire population.

\item We calculated the optical TDE luminosity function and we find a steep power-law relation $dN/dL_{g} \propto {L_{g}}^{-2.36}$, and a lower limit for the TDE rate of $\dot{N}\sim2\times 10^{-8}$ Mpc$^{-3}$ yr$^{-1}$, both consistent with previous measurements.

\end{enumerate}

The upcoming time-domain survey of the Rubin Observatory (LSST; \citealt{Ivezic2019}) is expected to observe 3500 to 8000 TDEs per year, depending on the assumed SMBH mass distribution \citep{Bricman2019}. In this work we showed that faint TDEs should constitute up to 50 - 60 \% of the entire population of these exotic transients. 
In addition, it is possible that there are  fainter TDEs but their population is hard to detect, especially in bright galaxies. 
Although technically difficult, we encourage the classification and follow-up of fainter nuclear transients in order to better understand the TDE rates and luminosity function. 
%than what the typical peak absolute magnitudes of TDEs is considered to be ($M\sim-(19-20)$~mag). In this way we can start building a larger sample of faint and fast TDEs and better understand the physical reasons behind their properties.

\begin{acknowledgements}

 We kindly thank D.B. Malesani for his help with Swift triggering and valuable comments and C. Angus for sharing the bolometric light curves of AT~2020neh. We thank the anonymous referee for comments that helped improve this paper. P.C, M.P and G.L are supported by a research grant (19054) from VILLUM FONDEN. P.C. acknowledges support via an Academy of Finland grant (340613; P.I. R. Kotak). IA is a CIFAR Azrieli Global Scholar in the Gravity and the Extreme Universe Program and acknowledges support from that program, from the European Research Council (ERC) under the European Union’s Horizon 2020 research and innovation program (grant agreement number 852097), from the Israel Science Foundation (grant number 2752/19), from the United States - Israel Binational Science Foundation (BSF), and from the Israeli Council for Higher Education Alon Fellowship. This work is based on observations made with the Nordic Optical Telescope, owned in collaboration by the University of Turku and Aarhus University, and operated jointly by Aarhus University, the University of Turku and the University of Oslo, representing Denmark, Finland and Norway, the University of Iceland and Stockholm University at the Observatorio del Roque de los Muchachos, La Palma, Spain, of the Instituto de Astrofisica de Canarias. This work used data taken with the Campo Imperatore Station as part of the first commissioning phase of Schmidt telescope's upgrade. The Campo Imperatore Station is part of the INAF-Astronomical Observatory of Abruzzo. We thank the observers Francesca Onori, Roberta Carini and Fiore De Luise.

\end{acknowledgements}

%-------------------------------------------------------------------
\bibliographystyle{aa}
\bibliography{bib.bib}

%%%%%%%%%%%%%%%%%%%%%%%%%%%%%%%%%%%%%%%%%%%%%%%%%%
%%%%%%%%%%%%%%%%% APPENDICES %%%%%%%%%%%%%%%%%%%%%
\appendix{}
\twocolumn
\centering

\section{Photometry tables} \label{apdx:phot_tab}

\begin{table}[h]
    \def\arraystretch{1.1}%
    \setlength\tabcolsep{6pt}
    \centering
    \fontsize{9}{11}\selectfont
    \caption{ZTF photometry in the AB system. All reported magnitudes are host-subtracted  and de-reddened.}
   
    \begin{tabular}{c c c c c}
    \hline
    \hline
        Date & MJD      & Phase (d)     & Band  &       Mag (Error)     \\
    \hline
2020-10-08   & $59130.5$ & $-20.9$ & $g  $ & $19.73$ ($0.24$) \\
2020-10-13   & $59135.4$ & $-16.1$ & $g  $ & $19.29$ ($0.12$) \\
2020-10-13   & $59135.5$ & $-16.1$ & $r  $ & $19.72$ ($0.19$) \\
2020-10-15   & $59137.5$ & $-14.2$ & $g  $ & $18.91$ ($0.10$) \\
2020-10-15   & $59137.5$ & $-14.1$ & $r  $ & $19.20$ ($0.13$) \\
2020-10-17   & $59139.4$ & $-12.3$ & $g  $ & $18.78$ ($0.11$) \\
2020-10-17   & $59139.5$ & $-12.2$ & $r  $ & $19.29$ ($0.15$) \\
2020-10-19   & $59141.5$ & $-10.3$ & $g  $ & $18.90$ ($0.09$) \\
2020-10-19   & $59141.5$ & $-10.2$ & $r  $ & $19.25$ ($0.14$) \\
2020-10-21   & $59143.4$ & $ -8.3$ & $g  $ & $18.45$ ($0.07$) \\
2020-10-21   & $59143.5$ & $ -8.3$ & $r  $ & $18.79$ ($0.07$) \\
2020-10-23   & $59145.5$ & $ -6.4$ & $g  $ & $18.15$ ($0.07$) \\
2020-10-23   & $59145.5$ & $ -6.3$ & $r  $ & $18.52$ ($0.07$) \\
2020-10-27   & $59149.4$ & $ -2.5$ & $r  $ & $18.16$ ($0.08$) \\
2020-10-27   & $59149.5$ & $ -2.4$ & $g  $ & $17.90$ ($0.06$) \\
2020-10-29   & $59151.4$ & $ -0.6$ & $g  $ & $17.78$ ($0.08$) \\
2020-10-29   & $59151.4$ & $ -0.5$ & $r  $ & $18.02$ ($0.06$) \\
2020-10-31   & $59153.5$ & $  1.5$ & $r  $ & $18.01$ ($0.06$) \\
2020-10-31   & $59153.5$ & $  1.5$ & $g  $ & $17.81$ ($0.07$) \\
2020-11-03   & $59156.5$ & $  4.4$ & $g  $ & $18.06$ ($0.07$) \\
2020-11-03   & $59156.5$ & $  4.4$ & $r  $ & $18.15$ ($0.08$) \\
2020-11-05   & $59158.5$ & $  6.3$ & $g  $ & $18.07$ ($0.09$) \\
2020-11-05   & $59158.5$ & $  6.3$ & $r  $ & $18.32$ ($0.08$) \\
2020-11-12   & $59165.4$ & $ 13.0$ & $g  $ & $18.84$ ($0.07$) \\
2020-11-12   & $59165.4$ & $ 13.1$ & $r  $ & $18.83$ ($0.09$) \\
2020-11-14   & $59167.5$ & $ 15.1$ & $r  $ & $19.32$ ($0.11$) \\
2020-11-14   & $59167.5$ & $ 15.1$ & $g  $ & $19.35$ ($0.11$) \\
2020-11-16   & $59169.5$ & $ 17.0$ & $g  $ & $19.85$ ($0.17$) \\
2020-11-16   & $59169.5$ & $ 17.1$ & $r  $ & $19.55$ ($0.14$) \\
2020-11-23   & $59176.5$ & $ 23.8$ & $g  $ & $20.00$ ($0.15$) \\
    \hline
    \hline
    \end{tabular}

\label{tab:ZTF_phot}
\end{table}

\begin{table}[h]
    \def\arraystretch{1.1}%
    \setlength\tabcolsep{6pt}
    \centering
    \fontsize{9}{11}\selectfont
    \caption{LCO photometry in the AB system. All reported magnitudes are host-subtracted  and de-reddened.}
   
    \begin{tabular}{c c c c c}
    \hline
    \hline
        Date & MJD      & Phase (d)     & Band  &       Mag (Error)     \\
    \hline
2020-10-24   & $59146.4$ & $ -5.4$ & $B  $ & $18.00$ ($0.03$) \\
2020-10-24   & $59146.4$ & $ -5.4$ & $V  $ & $18.34$ ($0.03$) \\
2020-10-24   & $59146.4$ & $ -5.4$ & $r  $ & $18.16$ ($0.01$) \\
2020-10-24   & $59146.4$ & $ -5.4$ & $g  $ & $18.39$ ($0.02$) \\
2020-10-24   & $59146.4$ & $ -5.4$ & $i  $ & $18.70$ ($0.17$) \\
2020-11-02   & $59155.4$ & $  3.3$ & $B  $ & $17.77$ ($0.03$) \\
2020-11-02   & $59155.4$ & $  3.3$ & $V  $ & $18.07$ ($0.03$) \\
2020-11-02   & $59155.4$ & $  3.3$ & $r  $ & $17.87$ ($0.02$) \\
2020-11-02   & $59155.4$ & $  3.3$ & $g  $ & $18.13$ ($0.04$) \\
2020-11-02   & $59155.4$ & $  3.3$ & $i  $ & $18.28$ ($0.17$) \\
2020-11-06   & $59159.4$ & $  7.2$ & $V  $ & $18.38$ ($0.04$) \\
2020-11-06   & $59159.4$ & $  7.2$ & $B  $ & $18.09$ ($0.03$) \\
2020-11-06   & $59159.4$ & $  7.2$ & $i  $ & $18.83$ ($0.20$) \\
2020-11-10   & $59163.4$ & $ 11.1$ & $B  $ & $18.57$ ($0.04$) \\
2020-11-10   & $59163.4$ & $ 11.1$ & $V  $ & $18.87$ ($0.05$) \\
2020-11-10   & $59163.4$ & $ 11.1$ & $r  $ & $18.74$ ($0.02$) \\
2020-11-10   & $59163.4$ & $ 11.1$ & $g  $ & $18.68$ ($0.04$) \\
2020-11-10   & $59163.4$ & $ 11.1$ & $i  $ & $18.80$ ($0.17$) \\
2020-11-17   & $59170.4$ & $ 17.9$ & $B  $ & $19.94$ ($0.09$) \\
2020-11-17   & $59170.4$ & $ 17.9$ & $V  $ & $20.24$ ($0.12$) \\
2020-11-17   & $59170.4$ & $ 17.9$ & $r  $ & $19.96$ ($0.04$) \\
2020-11-17   & $59170.4$ & $ 17.9$ & $g  $ & $19.75$ ($0.10$) \\
2020-11-24   & $59177.4$ & $ 24.8$ & $B  $ & $20.11$ ($0.13$) \\
2020-11-24   & $59177.4$ & $ 24.8$ & $V  $ & $20.19$ ($0.14$) \\
2020-11-24   & $59177.5$ & $ 24.8$ & $r  $ & $20.22$ ($0.06$) \\
2020-11-24   & $59177.5$ & $ 24.8$ & $g  $ & $20.06$ ($0.14$) \\
2020-12-06   & $59189.9$ & $ 36.9$ & $V  $ & $20.42$ ($0.17$) \\
2020-12-06   & $59189.9$ & $ 36.9$ & $g  $ & $20.44$ ($0.21$) \\
2020-12-07   & $59190.1$ & $ 37.1$ & $r  $ & $21.08$ ($0.13$) \\
2020-12-07   & $59190.4$ & $ 37.4$ & $B  $ & $21.29$ ($0.31$) \\
2020-12-13   & $59196.3$ & $ 43.1$ & $r  $ & $21.53$ ($0.24$) \\
2020-12-13   & $59196.3$ & $ 43.1$ & $g  $ & $20.66$ ($0.23$) \\
2020-12-21   & $59204.3$ & $ 50.9$ & $B  $ & $21.72$ ($0.52$) \\
2020-12-21   & $59204.3$ & $ 50.9$ & $V  $ & $20.81$ ($0.26$) \\
2020-12-21   & $59204.3$ & $ 50.9$ & $g  $ & $20.96$ ($0.34$) \\
    \hline
    \hline
    \end{tabular}

\label{tab:LCO_phot}
\end{table}

\begin{table*}[h]
    \def\arraystretch{1.1}%
    \setlength\tabcolsep{6pt}
    \centering
    \fontsize{9}{11}\selectfont
    \caption{Campo Imperatore photometry in the AB system. All reported magnitudes are host-subtracted  and de-reddened.}
   
    \begin{tabular}{c c c c c}
    \hline
    \hline
        Date & MJD      & Phase (d)     & Band  &       Mag (Error)     \\
    \hline
2020-10-29   & $59151.2$ & $ -0.8$ & $g  $ & $17.68$ ($0.18$) \\
2020-10-29   & $59151.2$ & $ -0.8$ & $r  $ & $17.07$ ($0.10$) \\
2020-11-10   & $59163.1$ & $ 11.1$ & $g  $ & $18.54$ ($0.10$) \\
2020-11-10   & $59163.1$ & $ 11.1$ & $r  $ & $19.05$ ($0.12$) \\
    \hline
    \hline
    \end{tabular}

\label{tab:CI_phot}
\end{table*}

\begin{table*}[h]
    \def\arraystretch{1.1}%
    \setlength\tabcolsep{6pt}
    \centering
    \fontsize{9}{11}\selectfont
    \caption{SWIFT photometry in the AB system. All reported magnitudes are host-subtracted  and de-reddened.}
   
    \begin{tabular}{c c c c c}
    \hline
    \hline
        Date & MJD      & Phase (d)     & Band  &       Mag (Error)     \\
    \hline
2020-10-28   & $59150.2$ & $ -1.8$ & $B  $ & $17.69$ ($0.16$) \\
2020-10-28   & $59150.2$ & $ -1.8$ & $U  $ & $17.58$ ($0.07$) \\
2020-10-28   & $59150.2$ & $ -1.8$ & $W2 $ & $16.95$ ($0.03$) \\
2020-10-28   & $59150.2$ & $ -1.8$ & $M2 $ & $17.24$ ($0.04$) \\
2020-10-28   & $59150.2$ & $ -1.8$ & $W1 $ & $17.42$ ($0.05$) \\
2020-11-01   & $59154.0$ & $  2.0$ & $W1 $ & $17.58$ ($0.04$) \\
2020-11-01   & $59154.0$ & $  2.0$ & $B  $ & $18.14$ ($0.17$) \\
2020-11-01   & $59154.0$ & $  2.0$ & $U  $ & $17.62$ ($0.05$) \\
2020-11-01   & $59154.0$ & $  2.0$ & $W2 $ & $17.35$ ($0.03$) \\
2020-11-01   & $59154.0$ & $  2.0$ & $M2 $ & $17.50$ ($0.04$) \\
2020-11-05   & $59158.0$ & $  5.8$ & $W1 $ & $17.79$ ($0.05$) \\
2020-11-05   & $59158.0$ & $  5.9$ & $M2 $ & $17.63$ ($0.04$) \\
2020-11-05   & $59158.0$ & $  5.9$ & $B  $ & $18.25$ ($0.18$) \\
2020-11-05   & $59158.0$ & $  5.9$ & $W2 $ & $17.51$ ($0.03$) \\
2020-11-05   & $59158.0$ & $  5.9$ & $U  $ & $17.93$ ($0.06$) \\
2020-11-09   & $59162.7$ & $ 10.4$ & $B  $ & $18.99$ ($0.32$) \\
2020-11-09   & $59162.7$ & $ 10.4$ & $W1 $ & $18.51$ ($0.06$) \\
2020-11-09   & $59162.7$ & $ 10.4$ & $U  $ & $18.45$ ($0.06$) \\
2020-11-09   & $59162.7$ & $ 10.4$ & $W2 $ & $18.45$ ($0.04$) \\
2020-11-09   & $59162.7$ & $ 10.4$ & $M2 $ & $18.46$ ($0.05$) \\
2020-11-13   & $59166.6$ & $ 14.2$ & $B  $ & $19.83$ ($0.70$) \\
2020-11-13   & $59166.6$ & $ 14.2$ & $U  $ & $19.20$ ($0.08$) \\
2020-11-13   & $59166.6$ & $ 14.2$ & $W1 $ & $19.15$ ($0.07$) \\
2020-11-13   & $59166.6$ & $ 14.2$ & $W2 $ & $19.05$ ($0.06$) \\
2020-11-13   & $59166.6$ & $ 14.2$ & $M2 $ & $19.14$ ($0.06$) \\
2020-11-27   & $59180.0$ & $ 27.3$ & $W1 $ & $20.02$ ($0.08$) \\
2020-11-27   & $59180.0$ & $ 27.3$ & $U  $ & $20.51$ ($0.08$) \\
2020-11-27   & $59180.0$ & $ 27.3$ & $W2 $ & $19.73$ ($0.06$) \\
2020-11-27   & $59180.0$ & $ 27.3$ & $B  $ & $20.97$ ($1.77$) \\
2020-11-27   & $59180.0$ & $ 27.3$ & $M2 $ & $19.81$ ($0.06$) \\
2020-12-04   & $59187.0$ & $ 34.1$ & $U  $ & $21.88$ ($0.13$) \\
2020-12-04   & $59187.0$ & $ 34.1$ & $W1 $ & $20.14$ ($0.13$) \\
2020-12-04   & $59187.0$ & $ 34.1$ & $M2 $ & $20.29$ ($0.19$) \\
2020-12-04   & $59187.0$ & $ 34.1$ & $W2 $ & $20.21$ ($0.12$) \\
2020-12-18   & $59201.4$ & $ 48.1$ & $W2 $ & $20.38$ ($0.14$) \\
2020-12-18   & $59201.4$ & $ 48.1$ & $W1 $ & $20.58$ ($0.18$) \\
2020-12-18   & $59201.4$ & $ 48.1$ & $M2 $ & $20.22$ ($0.16$) \\
    \hline
    \hline
    \end{tabular}

\label{tab:SWIFT_phot}
\end{table*}

\clearpage
\onecolumn
\centering

\section{\texttt{MOSFIT} posteriors} \label{apdx:graphs}

\begin{figure*}[h]
  \centering
  \includegraphics[width=19.5cm]{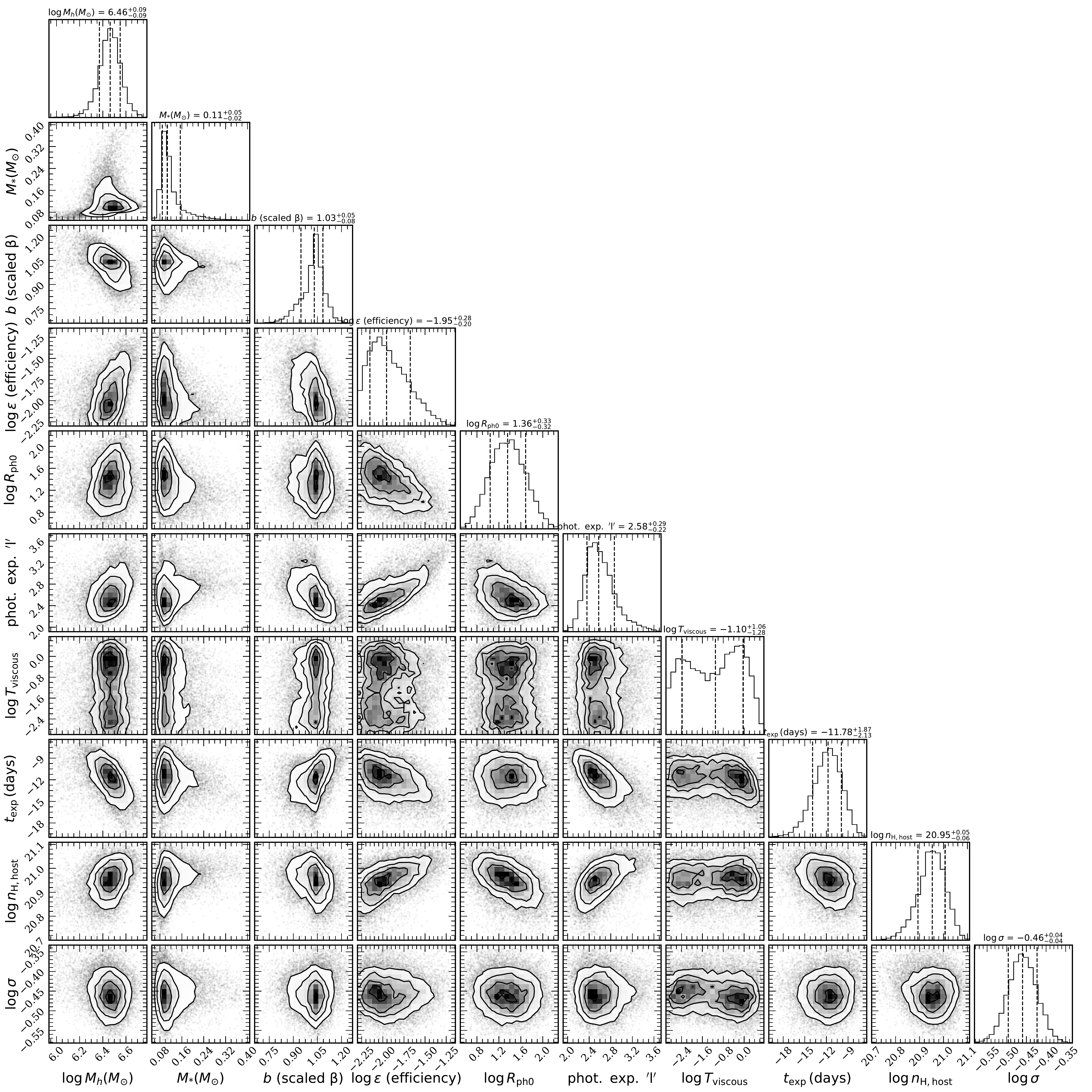}
  \caption{Posterior probability density functions for the free parameters of the model light curves in Fig. \ref{fig:mosfit}.}
  \label{fig:corner}
\end{figure*}

The priors and marginalized posteriors of our \texttt{MOSFIT} TDE model were listed in Table \ref{tab:mosfit}. In Fig. \ref{fig:corner} we plot the full two-dimensional posteriors.

\begin{figure*}
        \centering
        \includegraphics[width=0.5 \textwidth]{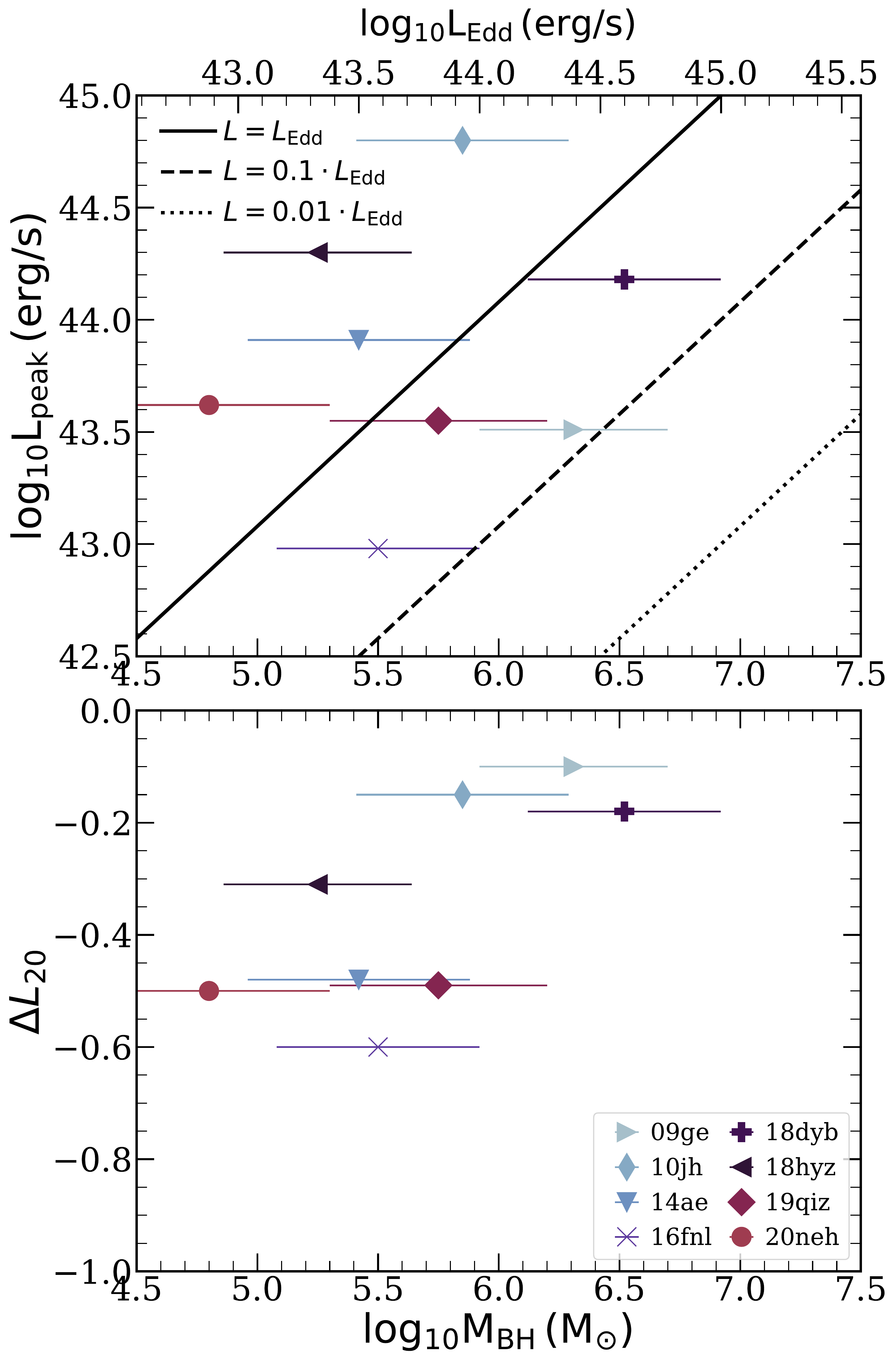}
        % \vskip\baselineskip
        \caption{Same as Fig. \ref{fig:Mbh_vs_L}, but for BH mass   estimated using the M--$\sigma$ scaling relation.  The correlation between the BH mass and the decline rate within the first 20 days ($\Delta L_{20}$) stands when   the M--$\sigma$ relation is used,  even though it is not as strong as the correlation detected in Fig. \ref{fig:Mbh_vs_L}.}
        \label{fig:Mbh_corr}
    \end{figure*}

\clearpage

% \onecolumn
% \centering
% \section{Tables}\label{apdx:tables}

\end{document}